# A CASE STUDY OF THE PICTURE ROCKS SUN DAGGER, PLUS A REVIEW OF THE INTENTIONALITY OF SUN DAGGERS

**Bradley E. Schaefer**
*Department of Physics and Astronomy, Louisiana State University,*
*Baton Rouge, Louisiana, 70803, USA*
Email: schaefer@lsu.edu

**James Stamm**
*11781 N. Joi Drive*
*Tucson, Arizona, 85737, USA*

**Abstract**: The Picture Rocks Sun Dagger is a spiral petroglyph on a hillside northwest of Tucson that shows sun dagger events at both the summer solstice and the equinoxes. On each of these dates, a wedge-shaped sunbeam with opening angles 20°–30° touches the center of the spiral, with both of these being confidently intentionally constructed by peoples of the Hohokam culture ca. AD 800–1300. More generally for claimed sun daggers throughout the American Southwest, the critical question is whether the ancient indigenous peoples *intentionally* placed the petroglyph so as to create a solar marker. The confident starting point for proving the intentionality of sun daggers in general is a histogram prepared by the Prestons showing highly significant peaks for indicated declinations within 2° of –23.4°, 0.0°, and +23.4°, with this being not by chance. In a review of solsticial and equinoctal sun daggers, we find that they all have beams of light shaped like a long-thin triangle with an apex opening angle of <40° that touches the center of the petroglyph symbol. While the majority of the sun daggers use a spiral petroglyph, circles and other symbols also are used. We find that from one-to-five light wedges appear on flat rock panels over a one hour interval of searching on just one side of a small hill, so false alarms must be common, and it is easy to find a place for a petroglyph so as to create an intentional sun dagger. Further, where a spiral or circular petroglyph has a coincidental light/shadow display, the false alarm rate is measured to be 20%–33%. Sun daggers that have indicated declinations other than ±23.4° or 0.0° are false alarms, including claims for alignments to cross-quarter days and lunar standstills, which are certainly wrong. Intentional sun daggers are not related to any form of calendric regulation, astronomical tools, or public ceremony. Rather, abundant ethnographic evidence shows that sun daggers are a part of sites, called Sun Shrines, where a local Sun-watcher would have lone vigils, with offerings and prayers to the gods on the solstices and equinoxes.

## 1 INTRODUCTION

A ‘sun dagger’ is a beam of sunlight that illuminates a petroglyph or pictograph in some significant manner on certain dates, particularly on the solstices and equinoxes. The prototype sun dagger is the one discovered near the top of Fajada Butte, in Chaco Canyon in 1979 (Sofaer et al., 1979). In this case, 49 minutes before local noon on the day of the summer solstice, a thin wedge-shaped beam of light was cast between some stone slabs to land exactly on the center of a ten-turn spiral petroglyph incised onto the cliff face behind the slabs. This sun dagger appeared on the spiral center for about 4 minutes on any date over a roughly two-week interval centered on the summer solstice. The striking coincidence of a wedge-shaped sunbeam covering the spiral center on the solstice was taken to be proof that the sun dagger was intentionally constructed by the Ancestral Pueblo culture as some sort of a calendrical or ceremonial marker. With wide publicity (e.g Carlson, 1983; Frazier, 1979; Solstice Project, 1982), this Fajada Butte Sun Dagger became iconic and widely recognized worldwide. The fame of Fajada Butte then inspired many searches and claims for more sun daggers throughout the American Southwest.

The sun dagger hypothesis has two severe problems. The first is that there is zero ethnographic support for considering anything even vaguely like the sun dagger idea, and there are zero analogies anywhere in the world (e.g., Zeilik, 1989). The second severe problem is that the American Southwest has a myriad of rock faces with shadows always shifting and all shapes being cast every day, while the Southwest also has a myriad of petroglyph symbols etched onto those rock faces, so we strongly expect a large number of random coincidences between the rock art and the light and shadows, even with zero knowledge or interest or intention by the ancient petroglyph-makers. Thus, at the start, the only evidence for the Fajada Butte Sun Dagger as an astronomical device was just the coincidence, while many such coincidences were expected with no intention on the part of the original petroglyph-maker. So in the early years, a reasonable and strong view was that the coincidence between the light wedge and the spiral center was just randomness (e.g., Carlson, 1983; Zeilik, 1985a). With no intention on the part of the builders, the sun daggers become trivial and uninteresting.

The evidence for/against *intention* with the Fajada Butte Sun Dagger changed dramatically in 1996 with the presentation of two papers at the Fifth Oxford Conference in Santa Fe. The two papers were finally published as Fountain (2005) and Preston and Preston (2005). Both papers presented statistical studies of many sun dagger sites throughout the American Southwest. Fountain reported on 45 sites from California to central New Mexico and from central Utah to southern Arizona. The Prestons reported on 46 sites, mostly different from Fountain's sites, mostly around central Arizona and New Mexico. The critical input was that these sun dagger sites mostly operated in the same way as the Fajada Butte Sun Dagger, with 37% pointing to spirals and circles, and two-thirds operating for the solstices and equinoxes. That is, the existence of many similar sites with the same provenance is very unlikely to be by chance alone (despite the many expected random coincidences), so the overall probability requires some causal connection that is essentially a proof of *intention* on the part of the sun dagger-makers. The Prestons' histogram (see Figure 1) of declination for interaction of a light beam with a petroglyph center shows strong peaks for the two solstices and a lesser peak for the equinox. The low-lying steady background level is caused by the many expected chance coincidences between the ubiquitous shadows interacting with the many petroglyphs, as these will point to random positions in the sky, with a continuum of indicated declinations. The equinoctal and solsticial peaks stick far above the background level, and are highly significant proofs that the majority of the sun daggers are not unintended random happenstance. In other words, the only way to get the peaks is for the makers of the petroglyphs to have intentionally placed them on rock faces to serve as something like seasonal markers. That is, most of the reported sun daggers are *intentional* constructs by the ancient petroglyph-makers.

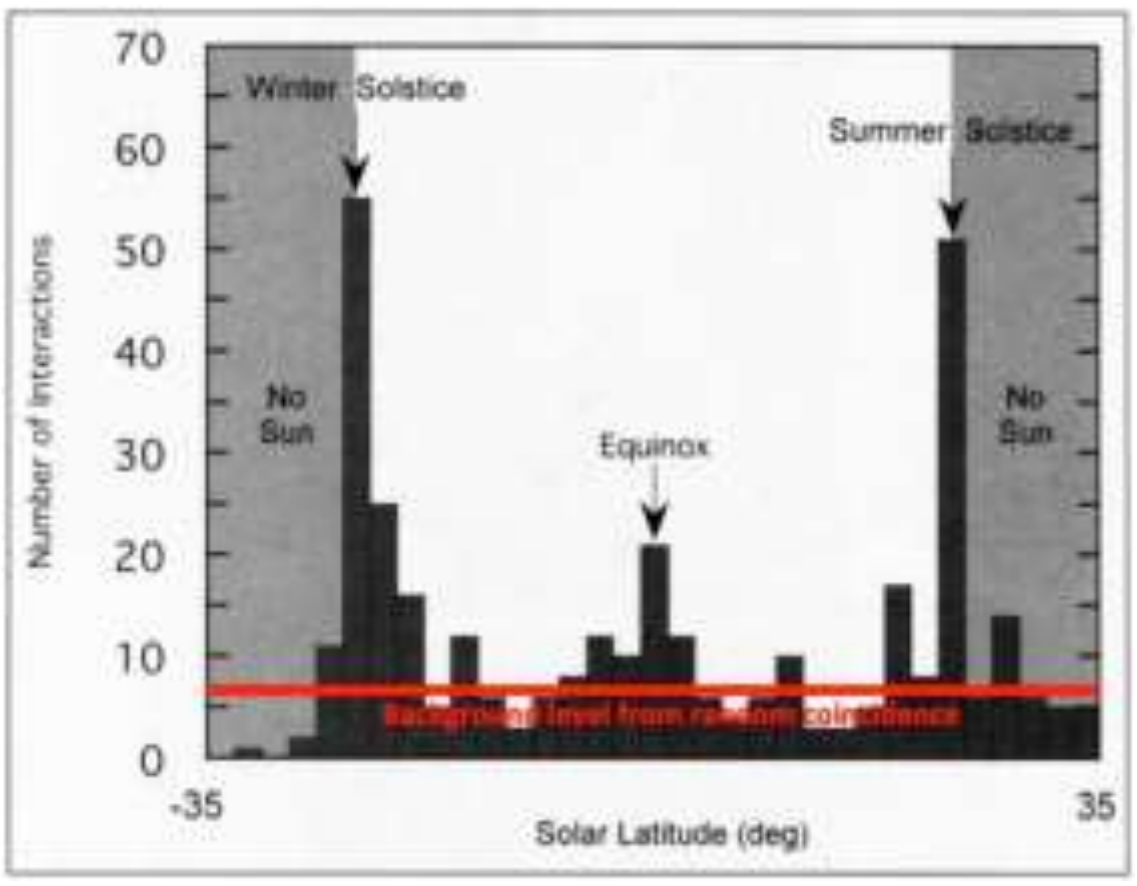

Figure 1: This histogram shows the distribution of indicated solar declination from potential sun dagger sites. The base figure is reproduced from Figure 7 of Preston and Preston (2005). One critical point from the Prestons' diagram is that there are highly significant peaks at the three declinations –23.4°, 0°, and +23.4°, corresponding to the solstices and equinoxes. This is the proof that many of the sun daggers were intentionally aligned by the ancient petroglyph-makers. The width of these three peaks demonstrates that the accuracy in construction is roughly 2° in declination, which is comparable to the stated measuring error. We also see that there is a background continuum level (at about 6.6) that is not focused on any particular declination, with this being the chance coincidence level. That is, the American Southwest has a myriad of petroglyphs on rock faces and random shadows will be cast on them all, such that many cases of apparent sun daggers can be 'discovered' despite zero intention on the part of the petroglyph-makers. This means that 'sun daggers' not pointing at either the solstices or equinoxes are random petroglyph/sunbeam coincidences with no intention and hence of no interest to anyone. Furthermore, some small fraction of sites apparently pointing at equinoxes and solstices are also random chance alignments. Comparing the area under the background level versus the area in the three significant peaks, the fraction of apparent sun daggers that are intentional is about 67%.

The essence for the proof of *intention* is that many cases of sun daggers with similar provenance and era all operate in the same manner (a wedge of light centered on a spiral or circle on the dates of the solstices or equinoxes). For this proof, we must have detailed case studies, allowing for real comparisons and statistics. Case studies will allow for defining the range of usages and properties as actually used by the early peoples of the American Southwest. Unfortunately, a good case study has been published for only the prototype on Fajada Butte. Other sun dagger sites have been published, but never more than a one-or-two pictures and a short description (e.g., Bates and Coffman, 2000; Houston and Simonia 2015; Zoll 2010). The two wonderful statistical papers by Fountain and the Prestons only aggregate some top-level facts, while specifically excluding site location information. So our scholarly community has a need for a second detailed case study of a sun dagger, and many more.

In this paper for the *Journal of Astronomical History and Heritage*, we present a detailed case study for one sun dagger close to Tucson, Arizona. Pamphlets and guided tours claim that this sun dagger operates on both solstices and on the equinoxes. This paper will give a full account of the observations of this sun dagger. Further, we give an analysis of sun dagger properties, with attention to evidences pointing to the intentions of the original creators.

## 2 PICTURE ROCKS SUN DAGGER SITE

This sun dagger is located 13 miles northwest of the center of Tucson, inside metropolitan Tucson, amongst low rock foothills on the northeastern side of the Tucson Mountains. The site is designated AZ AA:12:62(ASM), with the site name listed as "Picture Rocks" (AZSITE 2020). With the need for a real name (not just a string of forgettable letters and numbers), we are using the name "Picture Rocks Sun Dagger". Nevertheless, this name is problematic, partly because the site is not in the township of Picture Rocks far to the west, rather it is in metropolitan Tucson. The site name is also problematic because it is easily confused with the popularly-named "Picture Rocks" site on the edge of Picture Rocks township (also called Signal Hill) which also has a sun dagger, with the Picture Rock petroglyph site in western Arizona, and with the Picture Rock Pass Petroglyph site in Oregon. Further, the site name is confusingly similar to Painted Rock Petroglyph Site in central Arizona not far to the west of Tucson, and the Paint Rock Sun Dagger in central Texas. Despite these problems, we feel constrained to use the official site name as part of the sun dagger name.

The Picture Rocks site is on the Redemptorist Renewal Center property (a retreat owned by the Redemptorist Society, which is associated with the Catholic Church) with an address on Picture Rocks Road. Visitors are asked to check in at the main desk and to not climb on the rocks. Pamphlets are provided and signs point the way to the petroglyph panels. To the west of the main office building is a small rock hill, about 45 feet high, now with a large cross on the east side of the top, and a modern labyrinth constructed at the southern edge of the hill. A dry sandy-bottom wash runs past the western and southwestern sides of the hill, touching the cliff edges of the hill. The western half of the hill has many vertical and flat rock faces with many petroglyphs, including the Picture Rocks Sun Dagger (see Figure 2). The surrounding area is normally very dry, with flora and fauna typical of the Sonoran Desert.

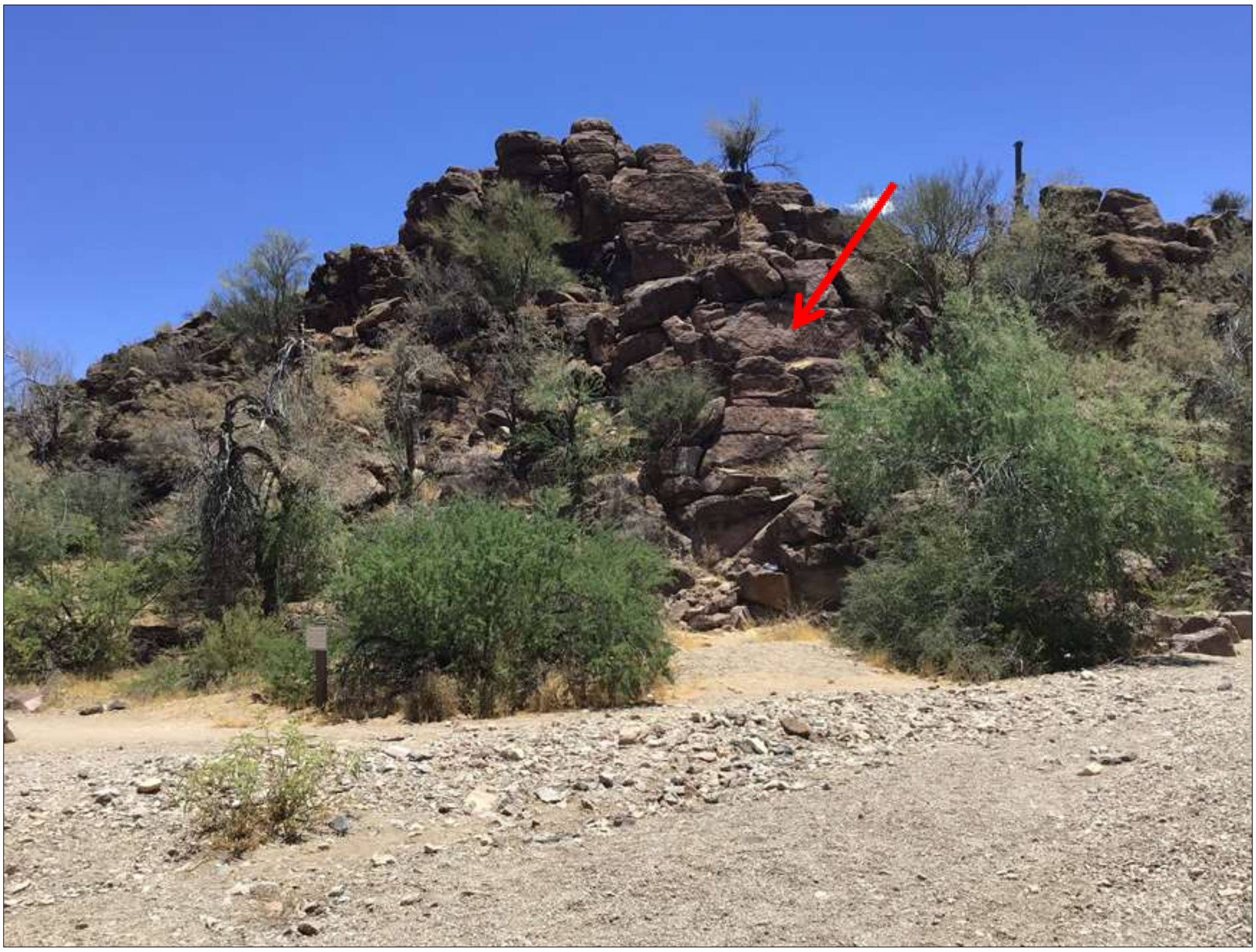

Figure 2: This picture shows the western side of the small hill on the Redemptorist property, with most of the local petroglyphs on this side. The location of the rock art panel with the spiral and the Picture Rocks Sun Dagger is identified with the red arrow (photograph: Brad Schaefer).

The locality has been surveyed by archaeologist Allen Dart, with a full site report given in Dart (2009). He found pre-contact Hohokam pottery sherds and flaked stone artifacts on the southeast, east and northwestern slopes of the hill. Other petroglyph sites are scattered throughout the Tucson Mountains and nearby areas, including Signal Hill (on the parklands of Saguaro National Park), 5.7 miles to the west-south-west. The larger area around Tucson has frequent archaeological evidence of the Hohokam culture, which flourished from ca. AD 300 to 1450. The Hohokam people are perhaps best known for their extensive irrigation canal systems and for their adobe four-storey 'Great House' at the Casa Grande Ruins National Monument. The Hohokam were sedentary, living in villages near water sources and arable land, and farming many crops, including maize, beans, squash, and cotton. The direct descendants of the Hohokam likely include members of the Akimel O'odham and Tohono O'odham nations.

Substantial details are known of Hohokam astronomy, both from archaeoastronomical studies and from ethnography with the Tohono O'odham (Bostwick, 2010). The cardinal orientations are prominent for ritual, ceremonial, architectural and cultural symbolism. The solsticial directions are prominent for the cosmology and for the sunrise calendar. Their calendar had a solar year along with 12 or 13 lunar months. The solar calendar was framed by the two solstices, with determinations of each solstice made by local specialist sky-watchers who looked at sunrise positions on the horizon and who watched sunlight cast through holes in buildings. The start of the New Year was centered on the harvesting of the saguaro fruit. The lunar months were named for local seasonal events, with the intercalation of the thirteenth month made irregularly from seasonal cues. The Hohokam had extensive lore and symbolism of at least the more prominent stars and constellations. This included the use of the Pleiades in five different phases for farming and hunting dates, although the defined times are vague enough that local seasonal events were likely to provide adjustments in real time. Year counts were not kept, but running calendar sticks with marks for important events (including meteor showers and eclipses) were made for the lifetimes of a few individuals. We have no hint that Hohokam astronomy went past the basics just described. Hohokam astronomy is identical in character (although all the names and stories are different) with that of the other populations in the American Southwest, and indeed with that of other indigenous cultures worldwide (Schaefer, 2017; 2018).

The Picture Rocks Sun Dagger is a spiral petroglyph on the west side of a small hill near the Redemptorist office (Dart, 2009). In a 1.5 hour reconnaissance in 2006, Dart counted 76 petroglyph panels on the top, south and east sides of the hill, while another 70 panels were on the west side of the hill (on the cliffs just above the wash), with some of the panels displaying more than one rock art design. These are all petroglyphs, made by hammering off the natural desert varnish on the surface of the rocks to expose the lighter-colored underlying rock surfaces. The rock art designs are the usual geometric patterns, animals and human stick figures. One of the most prominent and large panels, high on the western side of the hill, is shown in Figure 3. Critically, the spiral on this panel is that used for the Picture Rocks Sun Dagger. The archaeoastronomical potential for the spiral was apparently first recognized by non-indigenous people prior to 1997, as Rita Winters (2004) wrote in her book *The Green Desert: A Silent Retreat* that a woman from California who knew about the sun dagger showed it to Winters on the 21 June 1997, summer solstice day.

With permission, we climbed to the spiral petroglyph and closely examined it. The full horizontal width through the center is 47 cm, while the full vertical height is 39 cm. The spiral and much of its rock panel is on a light-colored surface, whiter and brighter than the lower edge of the rock panel that is covered with dark desert varnish. The spiral itself appears as an even lighter color surface, apparently having any remaining varnish pounded out by the spiral-maker. We saw no unusual coloration or contours or textures anywhere near the center of the spiral, indicating that any use of a marker to indicate the spiral origin has not survived the centuries. The light colored area has at least eight zoomorphic figures to the left of the spiral. The lower edge of the panel has eight stick figures holding hands such that it looks like a row of dancing people. The heads of the four leftmost figures have been knocked off, as a flake roughly two centimeters thick, exposing the underlying light-colored surface. To the upper right of the spiral, Dart has recognized a small protuberance of rock that has apparently been chipped off by human hands, with the break surface having the least varnish. Dart's discovery is important as it shows that the notch creating the equinoctal sun dagger was a man-made creation.

The timeline for the construction of the sun dagger can be given with relative milestones:

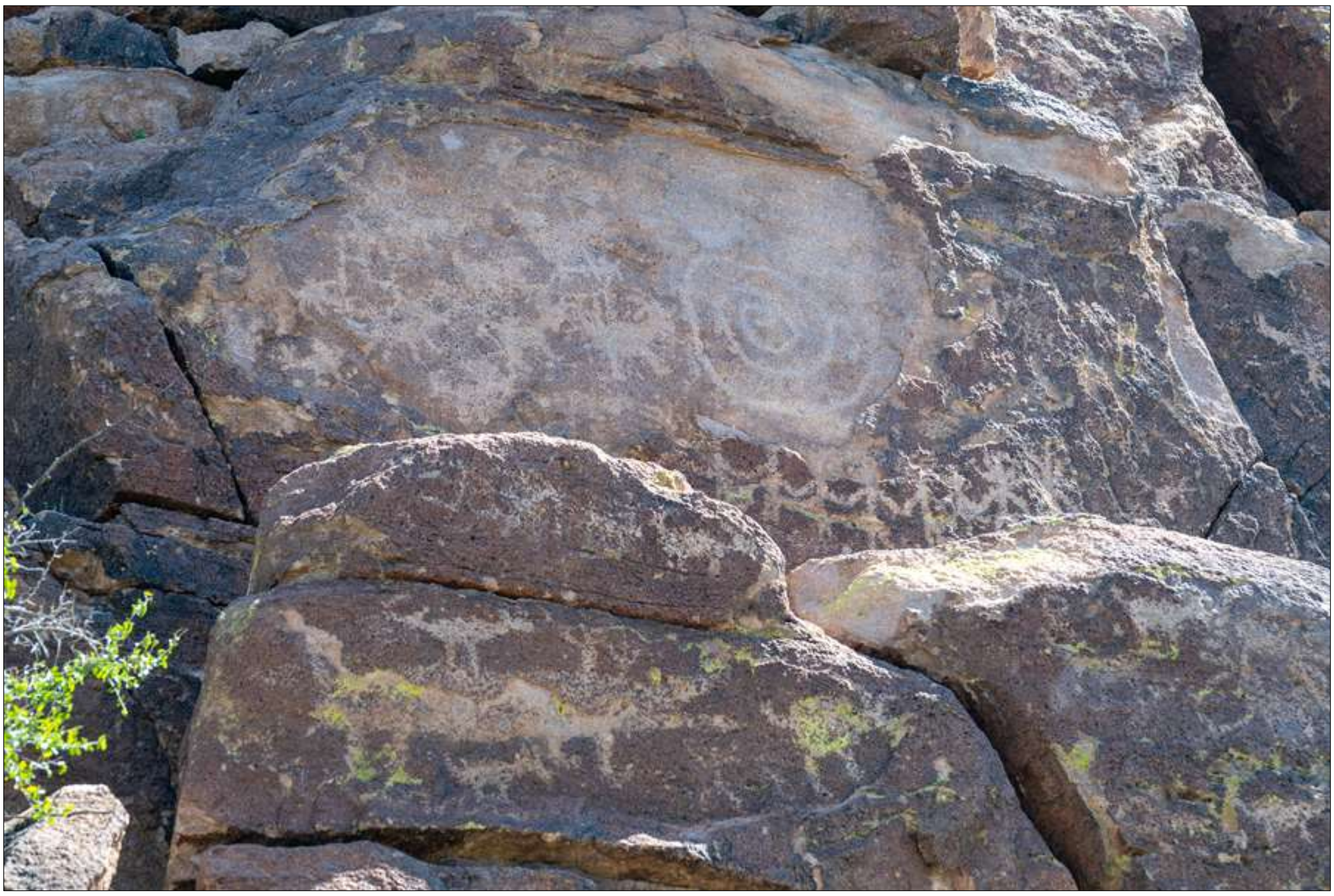

Figure 3: The rock art panel with the spiral that is the target of the Picture Rocks Sun Dagger. The spiral is easily visible (although not prominent) when in shadow, but the spiral becomes hard to spot when in full sunlight. Immediately below the spiral is a row of eight or nine stick figures holding hands, as if they are dancing together. See that the rock surface with the dancing men has been flaked off in parts, cutting off the heads of the four figures to the left, revealing a lighter colored surface on which the spiral appears. This would imply that the spiral is younger than the stick figures. To the left of the spiral, the panel shows about ten four-legged animals, some apparently with horns. The shadows cast to create the sun dagger are from minor protuberances in the rock face about a meter above the spiral and towards the upper-right of the spiral (photograph: Martha Schaefer).

(1) The original panel had heavy varnish, onto which the dancing-men and possibly other glyphs were made.

(2) In one or two flaking events of unknown origin, an unknown portion of the rock panel had the surface layer flaked off, removing perhaps up to one-inch thickness across at least the lower part of the rock panel. This flaking can be seen to have removed the heads of several of the dancing people. The uncovered rock is light colored and even now has little varnish.

(3) The sun dagger creators placed and pounded out the spiral and various animals on the left side. The spiral is of relatively low contrast due to the relative lack of varnish on its background surface. As we have found by frequent trials, the finding of a suitable position for a sun dagger petroglyph is but an easy task of looking around the desired area for an hour or less on the target date. With experiments with similar rocks at distant sites, we find that pecking or pounding to form symbols in the varnish is easy and relatively fast. The entire process of placing and creating a spiral sun dagger can easily be performed by one ordinary person in one day.

(4) The builders created the equinoctal sun dagger by chipping out a rock on the edge of the panel, thus creating an apparent notch that produced an elongated wedge of light.

The age of the spiral and the sun dagger is poorly known. Given the context of rock art in the Tucson area and the pre-European contact Hohokam artifacts found on-site, we can assign the petroglyphs and the spiral to the Hohokam, who occupied the area from ca. AD 300 to 1450. One of the petroglyphs apparently shows a bow and arrow, so that image is from after the introduction of the bow and arrow into southern Arizona either between 400 and 600 (Reed and Geib, 2013) or ca. 600 (Blitz, 1988). A large percentage of the glyphs on the hill are anthropomorphic and zoomorphic, which points to a date range of AD 800–1050 (Dart, pers. comm., June, 2020; Wright, 2014). The only stylistically-datable pottery sherd found at the site has been dated to AD 1000–1100 (Dart, pers. comm., June, 2020). None of the petroglyphs (other than the ob-vious modern graffiti) shows any signs of contact with settlers, so the panels are likely from earlier than 1500 or so. The use of spirals is common in Ancestral Pueblo (Anasazi) rock art, but only prior to 1300 (Schaafsma, 1979), so perhaps this Hohokam spiral is from before 1300. The spiral itself appears on a part of the rock face with relatively little varnish, having flaked off after the line of stick figures was incised (see Figure 3), so the spiral is not one of the earlier Hohokam symbols. In a local pamphlet handed out by the local Redemptorist Renewal Center, they summarize a reasonable conclusion that the Hohokam glyphs probably date to between AD 800 and 1300, while a more restricted date range of AD 800–1100 for the spiral is also reasonable (Dart, pers. comm., June, 2020).

This spiral is not the only spiral on the Redemptorist property. Intriguingly, Signal Hill (5.7 miles to the WSW) has 37 spirals all inside a 14 × 8 meter area on the top of a small rocky hill. The spirals in the Tucson area (and throughout the Southwest) show no preference for running outward in a clockwise or counterclockwise direction.

The petroglyph site has suffered a number of disturbances, as detailed in the site report (Dart, 2009). There are at least 8 instances of obvious modern graffiti, always short inscriptions with block letters and Arabic numerals. A 1928 newspaper article (Dart, 2009: 6) says:

> In the past few years a large portion of the valued rocks, which contain the writings of the early day residents of this vicinity, have been broken off, and in a number of cases carried away.

Dart (*ibid.*)

> … observed a couple of places near the bottom of the petroglyph hill on its west side, at about shoulder-level when standing in the wash bottom, where some of the hill's bedrock surfaces are much lighter than the surrounding desert-varnished rock, and in these lighter areas there are indentations or scars in the bedrock. This suggests that some chunks of the black-desert-varnished rock were removed at some time in the past.

Surface layers of loosely-attached rock sheets with high varnish are common on the hillside, so this flaking could be entirely natural.

## 3 LIGHT AND SHADOW ON THE PICTURE ROCKS SUN DAGGER

Our first visit to the Picture Rocks Sun Dagger was on 16 December 2019. On this date, the declination of the Sun ($\delta_{\odot}$) was –23.3°, practically identical to the winter solstice. On this date, first light near the spiral was at 10:41 a.m. Mountain Standard Time. (All times in this section are for MST, which is UT–7, because this part of Arizona never goes on Daylight Saving Time.) The entire panel appears to be all illuminated by 11:15 a.m. During this entire interval, tree branches partly shade the spiral panel, casting long-thin shadows that mildly wave back-and-forth. The interfering branches are from a Palo Verde tree growing out of the cliff about 20 feet to the south. Over time, the fractional coverage of the panel with tree-branch shadows is about 50%. Even with our video run at various speeds, we have difficulty in recognizing the boundaries of the shadow as it would be with no tree. At various times (e.g., 10:43 and 10:50 a.m.), a fraction of the shadows resemble a triangle that vaguely points towards the spiral center. These triangles of light are all seen to have their edges defined by tree branches. For times from 11:00 to 11:21 a.m., the spiral is dominated by long shadows variously moving due to solar motion and tree branch motion. None of these shadows looks much like anything we would take as a sun dagger shape.

On further visits from 21 December 2019 to 26 January 2020, we found the same situation with waving tree branches dominating the shadows on the panel. On the day of the winter solstice, at 11:00 a.m., a paid guided tour came and watched the shadows waving back and forth for roughly 15 minutes, while other tourists also looked for the solsticial shaft of light on the spiral. Throughout December and January, we were often not able to tell in real time that the tree-branch shadows dominated. With later viewing of sped-up videos from these days, we can realize that all the light and shadows were dominated by the tree branches. Further, even with the sped-up videos, we could make no good determination of the shadow positions for the case of no-tree. From this, we can make no confident determination as to whether a sun dagger would have been seen if the tree had not been in place. Still, no light/shadow display was seen at any time that could be described as any sort of a sun dagger.

With permission, we climbed the cliffs up to the spiral. With mirrors and cameras held close to the spiral center, we could see the 'skyline' of the rocks over which the Sun would rise on the spiral. With this, we can detect all possible sun daggers, independent of what is hidden by tree shadows. This skyline had two notches, which we label as the 'left-notch' and the 'right-notch', which will cast the summer sol-stice sun dagger and the equinox sun dagger respectively. For all positions below and south of the right-notch, where the Sun would appear to rise for all negative declinations, there are no notches or cracks or rocky protuberances. With this, we have a demonstration that there is certainly no sun dagger event of any description, intentional or unintentional, on the winter solstice or anytime from around the end of September up until the start of March.

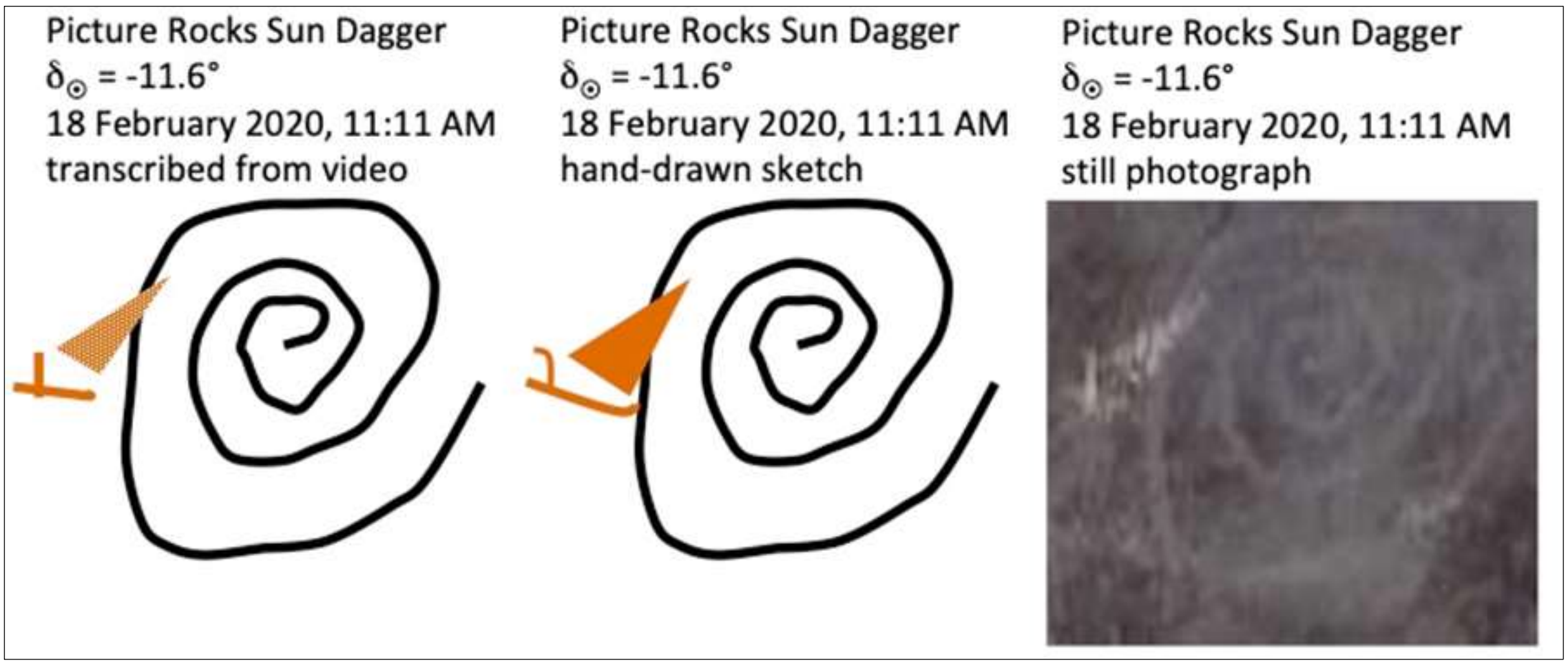

Figure 4: These images show three independent depictions of the sun dagger, as transcribed from our video, as copied from our hand-drawn sketch, and as taken as a still photograph. In this simple situation, we can see the imperfections and the complexities. The photographic image accurately shows the mottling structure with fine-detail, to the point where the structure appears relatively confused, and with the relative prominence of areas being distorted as compared to the appearance perceived by a human onlooker. The video record has poorer resolution than the still-photographs, but the videos have substantial advantages of coverage over every second of the entire event plus the ability to examine every fame at leisure. The hand-drawn sketches are necessarily simplifications of the complex sunbeams because the human artist only has a minute or so to draw in detail before the sunbeams change. The sketches have the big advantage of showing the light and shadow as seen by humans in real-time, with this being what would be seen by the Hohokam people. Fortunately, all of the input and conclusions in this paper are independent of the observing method or their usual uncertainties (photograph: Martha Schaefer).

Our data collection consists of video movies, some high-resolution photographs, plus hand-drawn sketches at typically two-minute intervals throughout. The video movies have continuous coverage in time, and we can stop and examine single instances of time with fine detail. The videos record the view with strong optical magnification, showing details past what ordinary humans could see. Further, the various lighting differences and such make for non-uniform single images, while the petroglyph is effectively invisible under bright sunlight. So we have transcribed the light and shadows onto a standard template for the spiral. We also have high-resolution still images on only a fraction of the days. The high magnification shows the complex mottling of light and dark (from slight divots and ripples in the rock) that are actually confusing for seeing the shape of the shadows. The hand-drawn sketches always have the shadow shapes simpler than is shown in the photographs, partly because there is too little time to draw all the fine details. The sketches show the view without all the invisible fine structure as would be provided with optical magnification, so the sketches provide a better representation of what a person would see. For a simple case, Figure 4 gives a near simultaneous rendition from all three recording methods. The times for all three images is close to 11:11 a.m. MST. As with all such images in this article, the title of each image indicates the date and time, plus the solar declination ($\delta_{\odot}$) at that time. The light beams on the panel are indicated with a burnt-orange color, with stippling to indicate either a relatively faint sunbeam or a mottled surface.

We visited the site on 18 February and 29 February 2020, with light beams as shown in Figures 4 and 5. On 18 February, some tree branch shadows were apparent towards the bottom, but these were of low density so that the shadow edges from the rocks were obvious. On both days, just after the time of sunrise on the panel, a wedge-shaped beam of light appeared on the upper-left side of the spiral. There was no apparent movement in the placement of the wedge between the two days, despite the declination of the Sun changing by 4.0°. This wedge could well be taken as an intentional sun dagger. However, the wedge was neither pointing at the center, nor pointing at the edge, of the spiral. On 18 February, the wedge started out as a sliver of light at 11:08, and had completely disappeared by 11:14 MST. Starting around this time, the panel received light at many places, all with complex shapes that we would not care to call a sun dagger. On 29 February, for just a few minutes around 11:20 a.m., the light beams coalesced into a region with the upper part forming a rough triangle with a tip half-way out inside the spiral. This wedge of light was 'sloppy' in shape and never pointed close to either the center or the edge of the spiral.

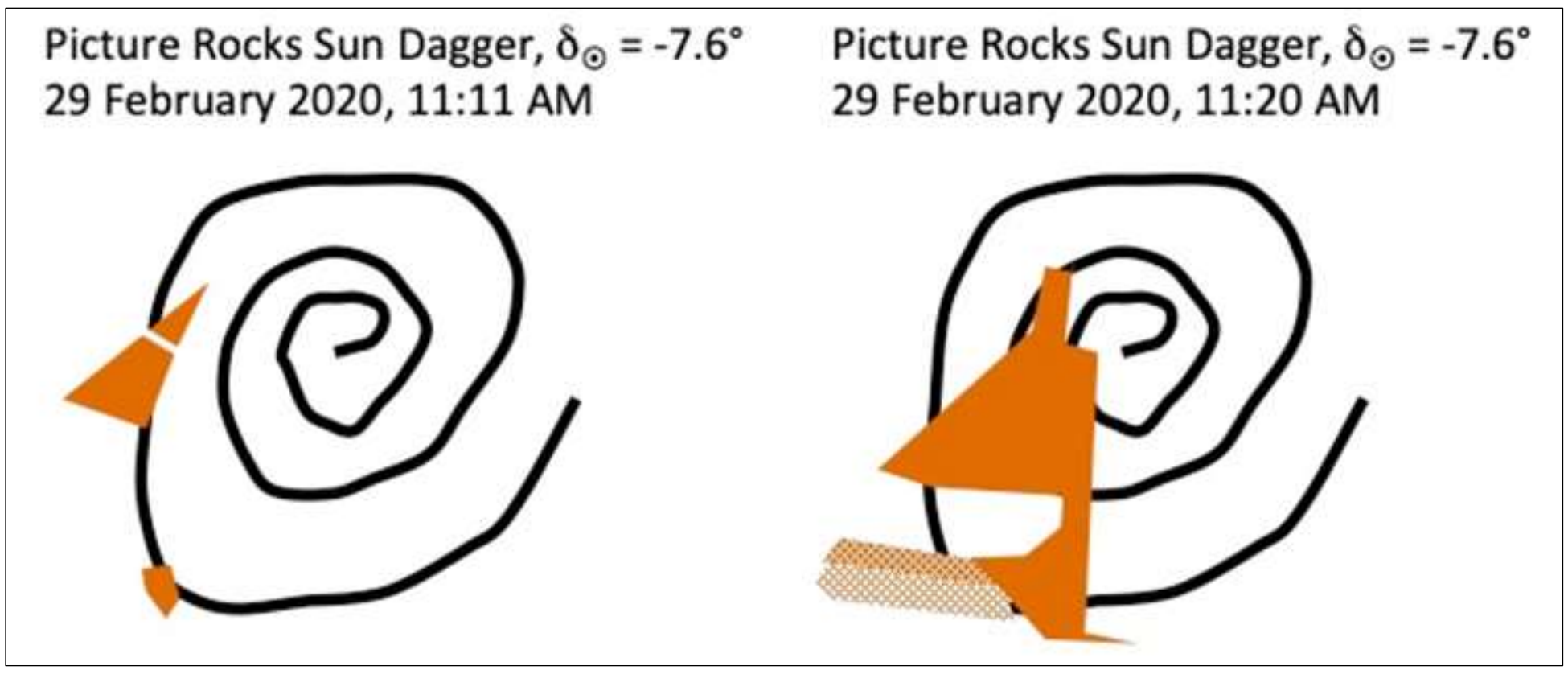

Figure 5: The sunbeams on 29 February appeared as roughly wedge-shaped patterns for two brief intervals centered on 11:11 a.m. and 11:20 a.m. The early wedge was just a continuation of the same wedge seen on 18 February. By itself, the early wedge was a simple and clear wedge-shaped sunbeam that might well be taken by a modern archaeoastronomer as an intentional sun dagger commemorating some date, perhaps labeled as Leap Day, with far-reaching calendrical implications. A 'zealous' archaeoastronomer could explain the miss on the spiral center as being deliberate, hence defining the radius of the spiral, and symbolizing the end of some calendar cycle. Alternatively, an 'enthusiastic' archaeoastronomer could take the sunbeam at 11:20 a.m. to be the intentional sun dagger, as it showed a good-enough wedge-shape and the sunbeam came near to the center. This shows the ease at finding false alarms and of making inappropriate interpretations. These false alarms illustrate the case where a true intentional sun dagger (in this case, for the equinoxes) will usually have sunbeams at other dates that can be taken as sun daggers (see the third method in Section 4). Such false alarms can be recognized variously because they do not have the classic wedge-shape, they do not interact with the spiral center, and they are not equinoctal/solsticial.

Our site visits on 20 March (the day with $\delta_{\odot}$ = 0°) and 24 March 2020 were aimed to see an equinoctal sun dagger and to see the change over just four days. The development of the light and shadow is shown in Figures 6 and 7. From 11:15 a.m. to 11:22 a.m., a wedge-shaped sunbeam formed to the lower right of the center, roughly pointed to near the center of the spiral. This sunbeam had a nice classic wedge-shape with an opening angle of around 30°, and did point at the center of the spiral, all with $\delta_{\odot}$ = 0°. At the end, just before the wedge disappeared by merger with other sunbeams, the apex of the triangle touched the center of the spiral. This was the equinoctal sun dagger.

This triangle-shaped sunbeam was formed by an apparent 'notch' in the rocks above the panel with a slight overhang, which we have called the 'right-notch'. This right-notch is what creates the wedges visible to the upper left of the spiral in February (see Figures 4 and 5). As the Sun moves north, the wedge is levered to the right. The right-notch is actually composed of two rock edges. The left side of the sun dagger wedge is cast by a rock slightly protruding above the rock panel and located 105 cm from the spiral center. The right side of the sun dagger wedge is cast by an edge of the rock panel and located 100 cm from the spiral center. A. Dart (pers. comm., June, 2020) has recognized since 2009 that the closer rock edge is formed by a projection that has been chipped away by a human hand. That is, the existence of the right-notch is human-made. Dart further realized that this modification of the notch so as to make a wedge-shaped sunbeam can only be intentional. That is, Dart's discovery and realization provides a demonstration that the equinoctal sun dagger at the Redemptorist property was intended by the ancient builders.

The Sun moved by 1.6° in declination between 20 and 24 March, so the shadows and the sun dagger had to move. From 29 February to 20 March (21 days), the wedge apex moved roughly two turns inside the spiral. So in four days, the apex should move roughly 40% of the turn-to-turn separation. Alternatively, for a 1.6° change in solar declination over a 100 cm length, the shadow should shift by 2.8 cm in these four days. But this day-to-day motion is confused by the minute-to-minute motion, so careful study of our videos showed no confident differences. For comparing 20 and 24 March as a real human observer, our sketches for 11:21 a.m. are indistinguishable. So even though the shadows should move over these four days in principle, the motion was not significantly visible in practice.

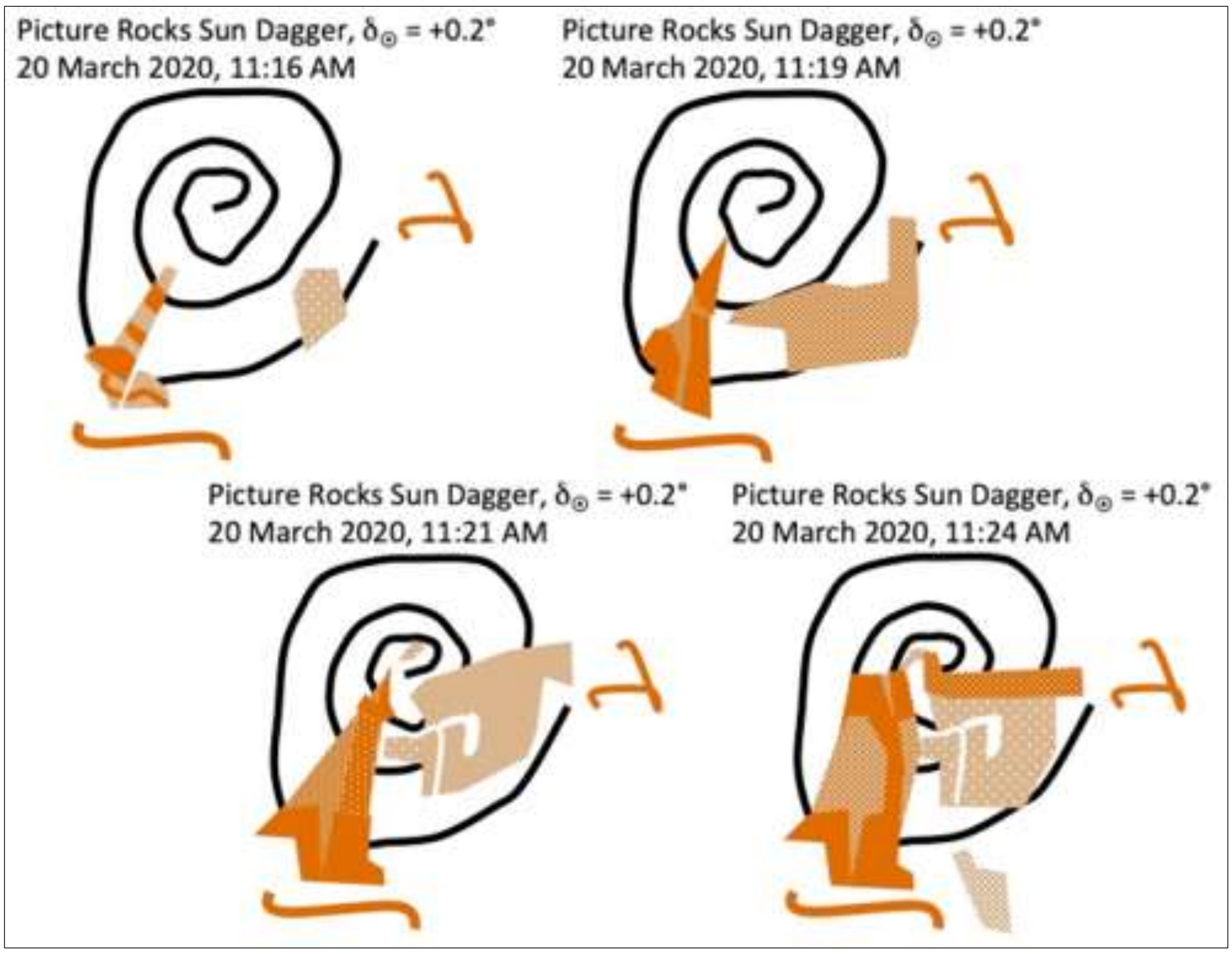

Figure 6: Sunbeams and shadows on the day of equinox, 20 March 2020. These transcriptions of the sunbeam patterns are from our series of hand-drawn sketches and the long-running video. From 11:15 a.m. to 11:22 a.m., a prominent wedge-shaped region appeared. The wedge pointed to the center of the spiral, and at 11:21 a.m. had its apex at the center of the spiral. This is a classic wedge-shaped sun dagger, covering the center of the spiral on the equinox.

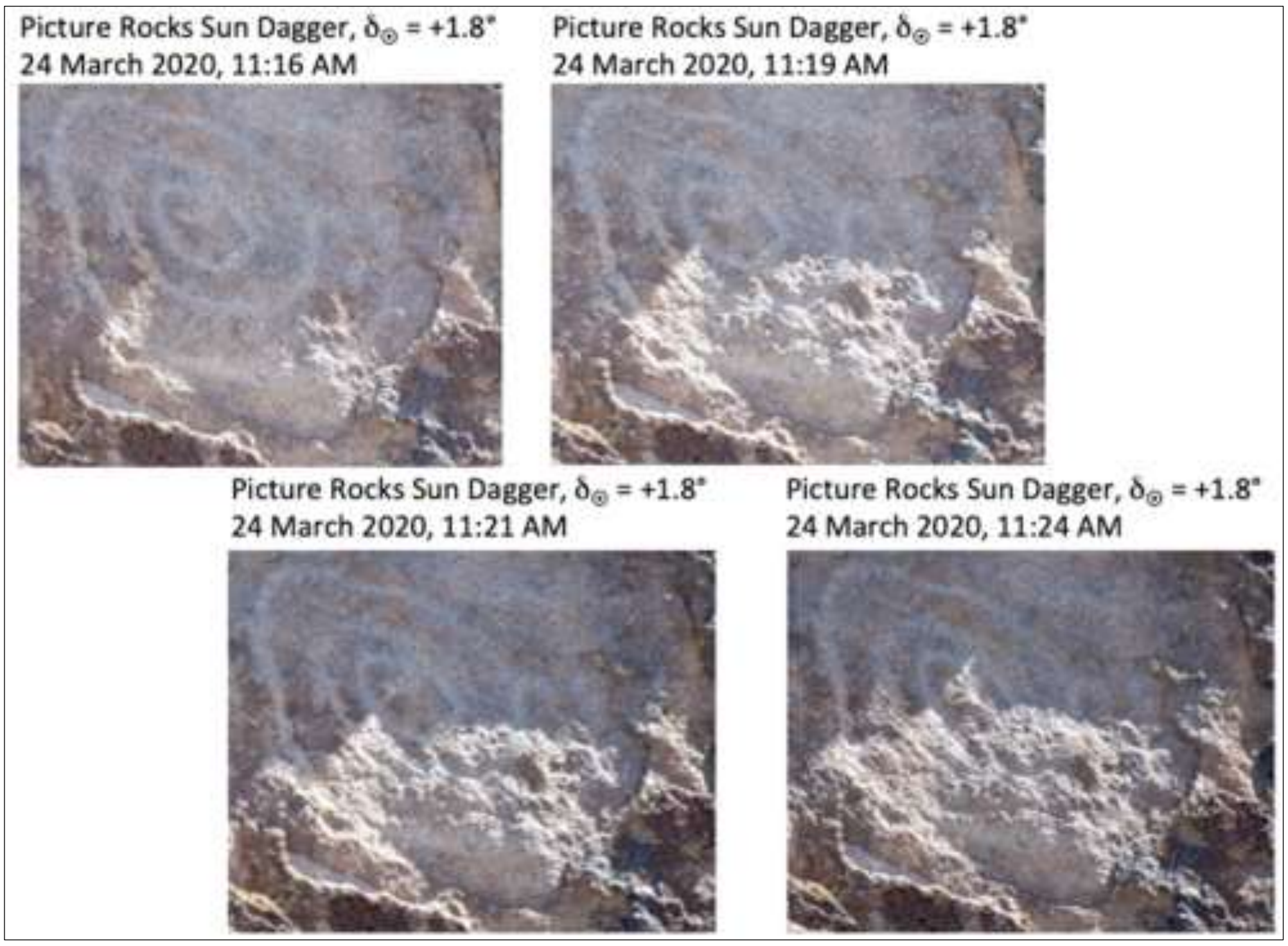

Figure 7: The sun dagger on 24 March 2020, four days after equinox. From 11:16 a.m. to 11:21 a.m., there was a prominent wedge-shaped light beam that pointed to near the center of the spiral. This was essentially the same light beam that formed the sun dagger on equinox day. Despite the Sun having moved north by 1.6° in these four days, the presentation was essentially identical to the human eye. Certainly, fine details had changed when looked at closely in photographs, but not by much, and the change in details was easily lost in the minute-to-minute changes. Critically, the hand-drawn sketches, which showed how the sunbeams were really perceived by humans in real time were indistinguishable from 20 March to 24 March (photographs: Martha Schaefer).

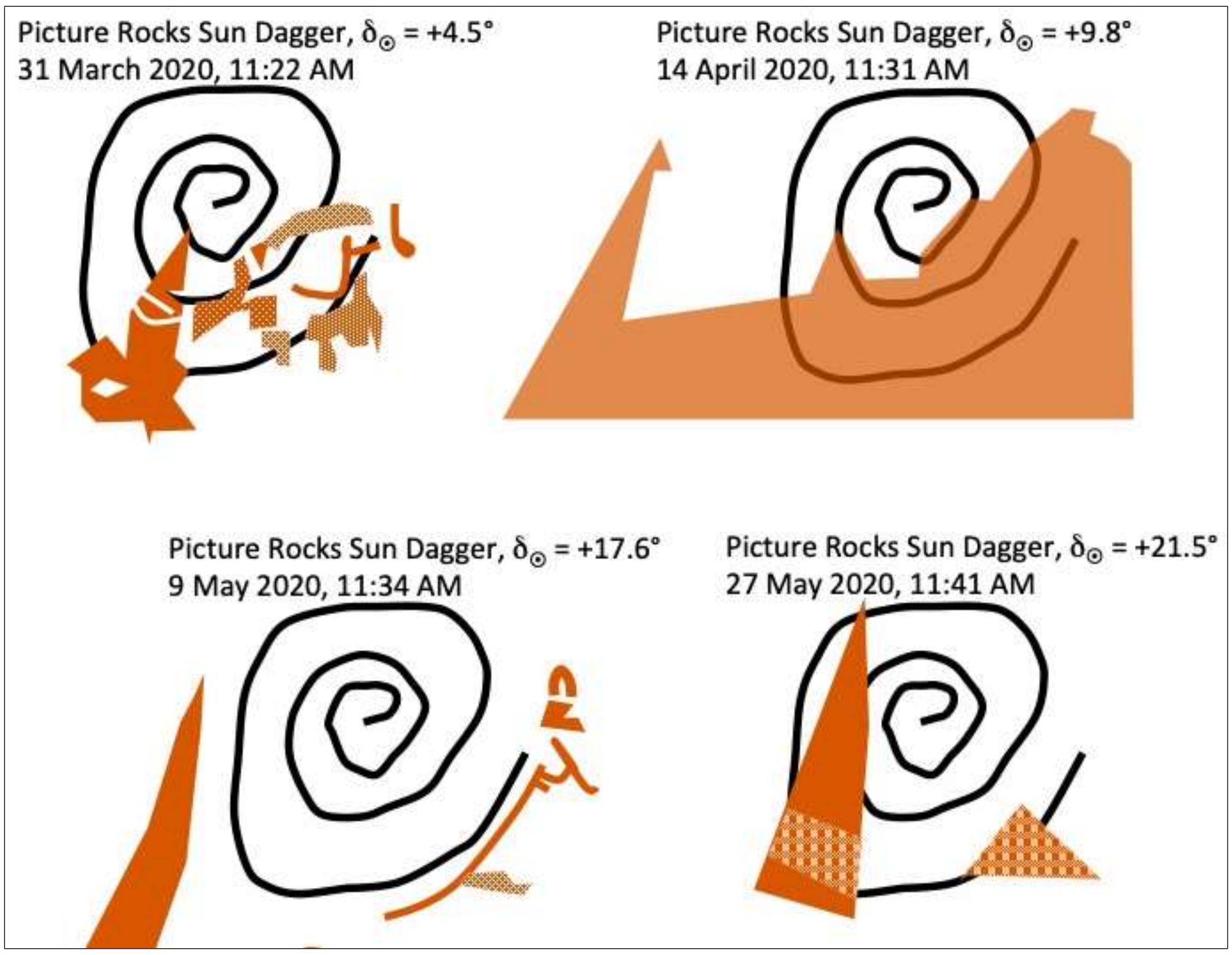

Figure 8: The sun dagger from 31 March to 27 May 2020, as the Sun's declination changed from +4.5° to +21.5°. The selected times for display are those for which the best wedge-shaped sunbeam was seen. Importantly, we see that every date from the equinox until the summer solstice has a wedge-shaped sunbeam that could be taken as a sun dagger by an enthusiastic archaeoastronomer. From 14 April until the summer solstice, the large vertical wedge on the left was steadily moving from left-to-right, with this wedge caused by a notch between two rocks jutting out as a slight overhang above the rock panel. As the Sun moved north, the position of the wedge was 'levered' to the right.

The further development of the light and shadows, from the equinox until the summer solstice, is shown in Figure 8. We see complicated shadows. But at each date, we can recognize wedge-shaped beams of light somewhere near to the spiral. If we were anxious to discover a sun dagger, these light-wedges on each date could be claimed to be significant, and we could find some justification to rationalize their positions on the rock art panel as being special and full of significance.

As the Sun moved north daily between winter solstice and summer solstice, the corresponding light wedge were levered towards the right. From 18 February until 31 March, we saw the wedge from the right-notch moving left-to-right across the rock art panel. After this, starting on 14 April and going until the summer solstice, we saw another wedge of light marching from left-to-right across the panel. This second notch will be named the 'left-notch'. The left-notch is a prominent gap between two boulders slightly overhanging the rock panel, about 135 centimeters from the spiral center.

Leading up to the summer solstice (20 June in 2020), the light wedge continued to move to the right. On the solstice, we saw a good wedge-shaped sunbeam, with opening angle of ~20°, shining on the spiral, with the right side touching the spiral's center in its later stages (see Figures 9 and 10). This was the classic form for a sun dagger. A high resolution time-lapse movie of this summer solstice sun dagger is presented in Schaefer et al. (2020), on the 5 August 2020 *Astronomy Picture of the Day*, which has now received over a million hits and been translated into 21 languages.

How much does the sun dagger wedge change in the days around the summer solstice? Going by the sketches, that is, as people see the sun dagger, the sketch for 11:49 a.m. on 20 June ($\delta_\odot$ = 23.43°) is indistinguishable from the sketch for 11:46 a.m. on 16 June ($\delta_\odot$ = 23.36°), and both are indistinguishable from the sketch for 11:42 a.m. on 27 May ($\delta_\odot$ = 21.46°). Thus, in practice, the sun dagger is unable to point out the date of the solstice to better than 24 days. This primary sun dagger is operating for at least 48 days each year. A further comparison can be made between videos from 20 June and 16 July ($\delta_\odot$ = 21.19°). Exact time synchronization is difficult, but it appears that the sunbeam triangle moved left by 3 cm over the 26-day interval when recorded at the same local solar time. But time-synchronized comparisons are not what the Hohokam would have done, rather all they could do is compare positions of the triangle when it passed some fiducial marker. For this, the triangle reached largely the same position (like touching the spiral center) on 16 July at a time approximately 3 minutes later than on 20 June. Thus, any observer would have a difficult time to distinguish the wedges on the two dates, even with time to closely examine magnified video images. So, again we conclude that the sun dagger motion was not significant for a visual observer over a 48 day interval centered on the solstice.

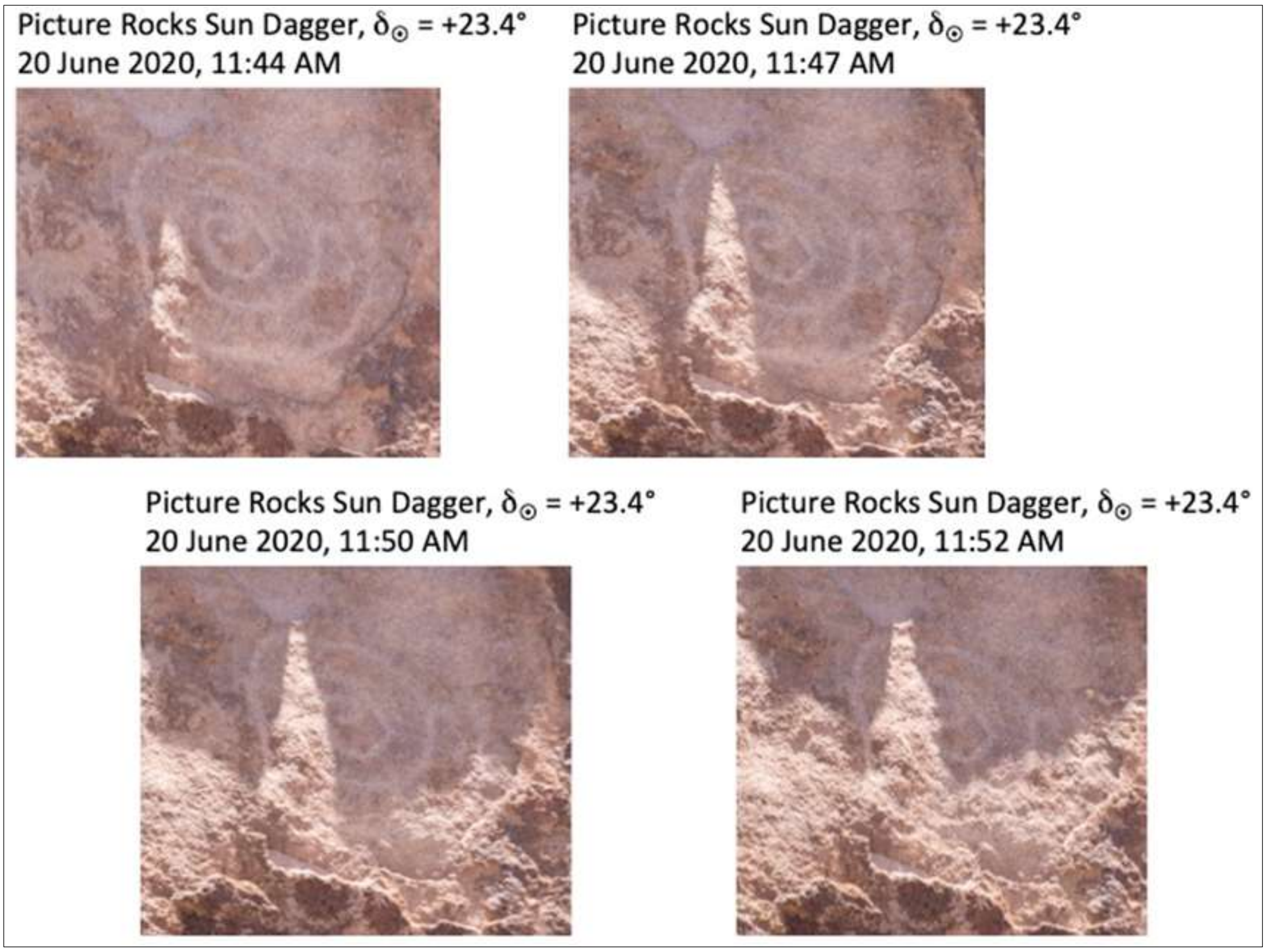

Figure 9: The development of the sun dagger on the day of the summer solstice. Around 11:47 a.m., the sunbeam took a classic wedge-shape, i.e., a long thin triangle with straight edges, with an apex opening angle of ~20°. By 11:50 a.m., the wedge expanded and shifted slightly to the right, so that the sunbeam was touching the center of the spiral (photographs: Martha Schaefer).

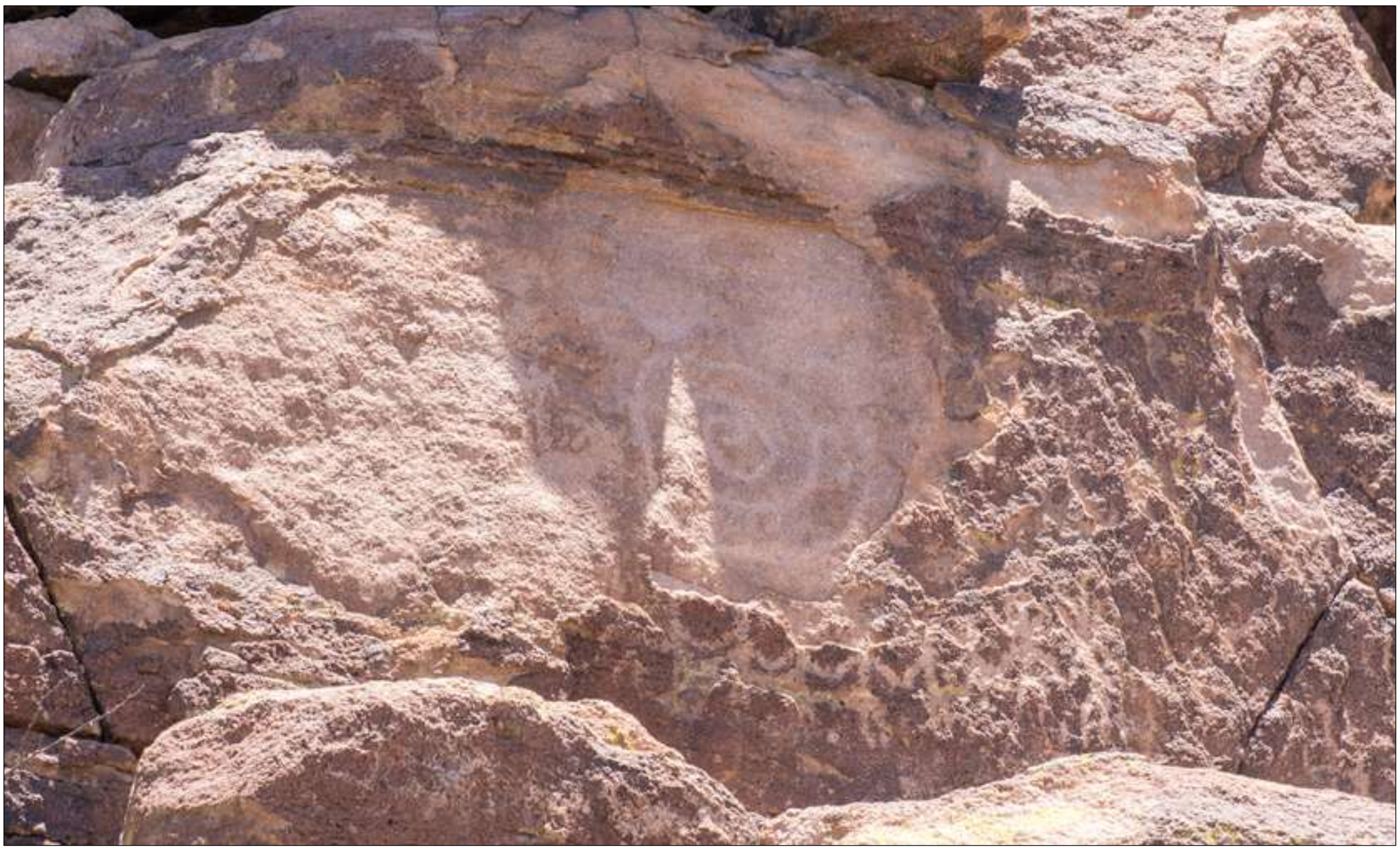

Figure 10: The summer solstice sun dagger at 11:47 a.m. on the day of the solstice. This shows the development of the classic wedge shape at the time when the wedge was the best triangular shape. Three minutes later, the wedge expanded and shifted to the right, so as to cover the center of the spiral (photograph: Martha Schaefer).

We have also looked closely at the sun dagger under the light of the Full Moon, looking for what we can call a 'moon dagger'. Before 'moonrise' on the panel, it was hard to recognize the correct rock panel on the darkened cliffside, and it was impossible to see the spiral. Moonrise on the panel was completely unobservable and unrecognizable. What is going on is that the very oblique illumination of the panel makes for very weak reflected light that is too faint to distinguish by the unaided human eye. The lunar illumination of the panel was first detected by the unaided eye at a time 30 minutes after moonrise on the panel. But by this time, any possible moon dagger event would have been long past. The illumination was still so weak that the edges of the shadow were too vague to recognize. Only 80 minutes after moonrise on the panel were edges recognizable in the shadows, and still with such low resolution that any sun dagger-like shape could not be recognized. All throughout, the spiral was never visible to the unaided eye. (And we have very long experience with such observational tasks, with our measurements showing that high experience improves detection thresholds by typically a factor of 3X. In addition, we were using five little-known techniques to improve visibility, including averted vision, frequent motion of the eyes, very-long dark adaptation, shading the Moon, and deep breathing.) Even with 10 × 50 binoculars, the spiral pattern was not seen at any time. The result of our observations is the sure knowledge that neither the spiral nor any wedge-shaped moonbeam could possibly have been visible at the Picture Rocks Sun Dagger at any night-time.

The frequency of wedge-shaped sunbeams is critical for knowing how hard it is for the Hohokam to find appropriate sun dagger sites on which to place a petroglyph, plus for getting some idea of the false alarm rate. To get some measure of the rate of wedge-shaped sunbeams on flat rock faces, we spent time looking for wedges of light appearing on the western half of the hill. We found that narrow triangles of light were common. With casual and brief looking from February to May, we could always find a good case where a petroglyph could be placed to create a sun dagger for that date. For a longer search done lackadaisically, from 11:00 a.m. to noon on 27 May, we found and photographed nine wedges of light. Two of the wedges have undetermined nature (because we could not climb on the rocks to get close), two others were formed by pairs of rock edges joined at an angle (so they would not move or readily have a spiral placed appropriately), and four wedges consisted of a rock edge intersecting a moving shadow (where a classic sun dagger cannot be created). One of the wedges was a simple and pure wedge shape (with a 20° angle) all displayed on a flat rock panel where a spiral could easily be placed to create a sun dagger. We made a serious search from 11:00 a.m. until noon on 16 June, during which we identified 5 classic wedge-shaped sunbeams edged by shadows onto flat rock panels where a petroglyph spiral could have been placed to create a solsticial sun dagger. That is, we found five sun dagger possibilities in just one hour on the side of one small hill. This demonstrates that finding appropriate light wedges is easy and common.

This section has reported in detail a full case study of our observations of the Picture Rocks Sun Dagger from before the winter solstice to after the summer solstice. This covers all possibilities, with solar declinations from –23.4° to +23.4°. The phenomena from the summer solstice to the winter solstice depends only on solar declination, and so are identical to that for which we already have good data. Thus, with only a bit more than six months of data, we have exhaustively recorded the sunlight phenomena associated with the Picture Rocks Sun Dagger.

## 4 WHAT IS THE FREQUENCY OF FALSE ALARMS?

'False alarm' is standard terminology throughout the sciences where measurements look to imply some discovery or effect that is actually casual (i.e., not causal). In the context of sun daggers, a false alarm would be some real sunbeam/petroglyph pattern that is random coincidence with no intention by the original petroglyph-makers. Without the intention of the original makers to create a sun dagger, the false alarm sunbeam/petroglyph pattern is just random noise with zero significance for the old cultures, and hence is of no interest.

We know that many petroglyphs throughout the American Southwest will randomly have shadows cast upon them that can be interpreted as a sun dagger, despite there being no knowledge or intention of this phenomenon by the petroglyph carvers. That is, a substantial false alarm rate must occur, where the apparent sun daggers are chance coincidences with no intention. Such false alarms are inevitable, of no interest for anyone, and cause confusion and noise for the real sun daggers. The critical question is "What is the frequency of the false alarms?"

Many workers have recognized the critical nature of the false alarm rate, but no one has ever quantified the rate. Here, we will make quantitative estimates in three different ways.

The first way is to ask about the fraction of possible or proposed sites that do not have intention. For this, we use the Preston histogram (Preston and Preston, 2005: Figure 7, or see our Figure 1). Each declination is for all possible light beam interactions from 60 spirals and 46 circles exhaustively taken from 14 sites in a small area in eastern Arizona. We see there are big peaks, each highly significant, at indicated declinations of ±23.4° and 0.0°, with a width of about 2°. Such a histogram could not occur by chance. There must be a causal mechanism operating to produce the three peaks. And this mechanism can only be for the many individuals who pecked the petroglyphs to somehow place the petroglyph based on the solsticial or equinoctal position of the Sun. This is the proof that the majority of the sun daggers are *intentional*. This is also a proof that the petroglyph-makers were aiming with reasonable accuracy towards the solstices and equinoxes. But what about the many sites with the indicated declinations between the three peaks? These are the false alarms. That is, there is no realistic way that the makers could have intentionally chosen to construct petroglyphs at locations such that they create a continuum of indicated solar declinations from –34° to +35°. There is a flat baseline level that constitutes the random coincidences (with the intentional peaks sticking out high above). This background noise level can be estimated by averaging all the values away from the three intentional peaks. This is 6.6 inside each bin in the histogram. This noise level must continue within the equinoctal/solsticial peaks, which is to say that some fraction of the sun daggers pointing at the equinoxes or solstices are not intentional. With this, 18% of apparent sun daggers pointing at solsticial/equinoctal declinations are random coincidences of shadows and petroglyphs. With this background level of 6.6 per bin, 33% of all spiral/circles indicate declinations that were not intended by the makers.

This same method can be used with the summary statistics in Fountain (2005). His Table 1 lists that 80% of the sun daggers in his sample are equinoctal or solsticial. These are the intentional sun daggers. This leaves 20% that Fountain says point to the cross-quarter positions. Section 5.7 below will demonstrate in multiple convincing ways that these 20% cannot be intentionally pointing to the Sun on these days, so they must be unintentional. Now, our two measures of the overall false alarm rate (i.e., 33% and 20%) are consistent, with the differences undoubtedly arising from ordinary sampling and selection differences. In all, a spiral or circle petroglyph, with shadows cast upon it, has between a ⅕ and a ⅓ chance of being an unintentional false alarm.

A second method to quantify the false alarm rate is to use our searches for sunbeam wedges on flat rock panels. Our searches were only for the western half of the one small hill on the Redemptorist property, yet such will be typical of rock piles and cliffs throughout the American Southwest. On every date, we found one to five classic wedge shape sunbeams move across some relatively flat rock panel. (This is not counting the wedges associated with the spiral of the Picture Rocks Sun Dagger.) If a petroglyph were carved near one of these random wedges, then an enthusiastic latter-day researcher could identify another sun dagger. Another way to say this is that an ancient person seeking to create an intentional sun dagger will easily be able to find an appropriate location to place a spiral (or other petroglyph). That is, if searching on one side of one small hill, typical of the region, produces 1–5 reasonable wedge-shaped sunbeams on a flat rock surface on all dates, then a sun dagger-maker can readily find some local position to carve the petroglyph such that an intentional sun dagger is created. There is nothing rare or special about sun dagger sites. Further, every hill and cliff can easily generate false alarms, at a rate of about 1–5 per hill/cliff.

A third method to quantify the false alarm rate is to look at the multiple sun daggers interacting with one petroglyph. That is, if a genuine intentional sun dagger is memorialized on one date, then are any sun dagger events on other dates false alarms? This is an important question for sun daggers that are claimed to work on multiple dates. The Fajada Butte Sun Dagger is claimed to work on four declinations; +23.4° (summer solstice), 0.0° (equinox), –23.4° (winter solstice), and +28.6° (northern lunar standstill). Were all these intentional, or are two or three of these just part of the false alarm rate?

For this type of false alarm, we have strong practical constraints. While it is easy to find positions on rock panels that satisfy one target declination, the probabilities are very small that the same spot will also satisfy a second target declination. For any given sun dagger to work on another date, the solar motion requires that a widely separated rock structure cast the critical sunbeam on the second date, but rock structures are uncorrelated for wedge-shadow-casting notches, so the probabilities are uncorrelated. Let us label this probability as "P". Crudely, the probability that a point on a rock surface with an operational sun dagger will also have a wedge-shaped sunbeam operating at that same point on some other target date is equal to the ratio of the area of the petroglyph to the area of flat rock panels all over the hillside (see method two above). With photos of our hillside, we estimate $P \sim 10^{-5}$. P will vary greatly from site-to-site, but an order-of-magnitude estimate is adequate for the discussion here. The point is that P is very small and incredibly unlikely, while $P^2$ is impossibly small. We could postulate that the original builders cast around on many hills and cliffs searching amongst many potential wedge-shaped sunbeams that work for one target date, somehow memorized, for just the very rare point that also works on some other target date. They would have to search ~1/P sites with a working sun dagger on one target date to see whether it also works on the second desired target date. With a huge and directed effort, such a site might be located. But such a program is not within ordinary cultural practice for the ancient peoples of the Southwest. Anyway, such a massive and implausible program is not needed because creating two separate sun daggers for the two target dates is so easy and just as good. So a single petroglyph with *two* sun dagger dates is *possible*, but very unlikely.

For a single sun dagger to work for three target dates, we have to square a very small probability. That is, the probability of finding one point on a single hillside that intentionally works on three separate dates is $P^2$, which must be incredibly small. Effectively, it is impossible to have one spot on a rock panel work as a sun dagger for three target declinations.

This stark reality can be circumvented if the builders modify the rocks, for example by knocking out some strategically placed notch. For the Fajada Butte Sun Dagger, the rock slabs and their edges are certified to be all natural (Newman et al., 1982). However, for the Picture Rocks Sun Dagger, the notch for the equinoctal sun dagger appears to be human-made. With this, the Picture Rocks Sun Dagger can have intentional sun daggers on two independent dates, with no problems with extreme probabilities.

The stark reality (i.e., the severe improbability of two or more intentional sun daggers for one petroglyph) can be moderated if the second and third target declinations are allowed to have their interactions far away from the original center or to have some shape other than wedge-shaped. That is, if we allow any shape of sunbeams interacting anywhere around the petroglyph to be recognized as a sun dagger, then the effective P value is made greatly larger, and multiple indicated declinations becomes much less improbable. The trouble is then that nearly all petroglyphs operating as a sun dagger on one date can operate as sun daggers at almost any other date. This is because a notch that casts a shadow on the first target date will cast a shadow elsewhere on the rock for any second and third target dates. A modern commentator can always invent some post-facto justification for any configuration of sunbeam and rock art, hence creating a fallacious rationalization for the second unintended sun dagger. In all, we are left with the reality that having two target declinations for the center of one petroglyph is very unlikely under natural circumstances (so at least one of the two targets is a false alarm), while having three or more means that the extra dates are certainly false alarms.

An application of this is to the Fajada Butte Sun Dagger, with four target declinations. Certainly two-or-more of those targets are false alarms. Most everyone takes the lunar standstill alignments to be random and unintentional (e.g., Ruggles, 2005; Zeilik,1985a, 1989), for many strong reasons. The equinoctal sun dagger actually operates off a second spiral located to the upper left of the famous spiral, with the narrow sunbeam knifing through the spiral's center. The use of a second spiral means that the above probability argument is not applicable for this claimed sun dagger. (The equinox also has a narrow wedge on the main spiral, but it is far off-center and looks to be a false alarm. This is just another product of the effect that one real centered sun dagger will make for an off-center sun dagger on other dates.) The winter solstice has two narrow wedges of light appearing at random positions inside and out-side the spiral, so there is nothing to distinguish these dates from any other dates, given that the daggers are always moving back and forth throughout the year. So the strong community opinion is that at most two sun daggers at Fajada Butte are intentional (ibid.). With this, two or more out of four claimed daggers at Fajada Butte are false alarms.

We know of only two other cases with adequate published data where possible sun daggers are indicated for two or more declinations. The Picture Rocks Sun Dagger has three claimed sun markers, for the winter solstice, the equinoxes, and the summer solstice. For the winter solstice, there is no wedge of light, or anything, so this sun dagger does not exist and we are down to two indicated declinations. For the summer solstice and equinoxes, wedge-shaped sunbeams touch the center. Both look to be good sun daggers. Critically, A. Dart (pers. comm., June, 2020) points to the man-made nature of the chipped right-notch for the equinoctal sun dagger, and this circumvents the strong probability argument for only one indicated date for any one petroglyph. So here is a case where one petroglyph confidently marks two solar declinations. Still, the Picture Rocks Sun Dagger has one claimed sun dagger event (for the winter solstice) that is a false alarm.

The other case of a double sun dagger is for the Hayden Butte spiral petroglyph, where the summer solstice has a 40° wedge—with its apex close to the center of the spiral, and where the winter solstice is marked by a rounded right-angle of shadow (Bostwick, 2010). The summer solstice event looks to be a classic sun dagger. The winter solstice sun dagger is not much of a wedge and is the only case published where the *shadow* is the pointer (not a sunbeam). So it looks like the summer solstice sun dagger is intentional, and the problematic winter solstice sun dagger is a false alarm.

Unfortunately, for calculating a quantitative false alarm rate, we only have three cases with explicit coverage. In particular, for other sun dagger sites, no one has published whether additional apparent wedges of light interact at other dates in the year. Given the three case studies available, it appears that all intentional sun daggers are accompanied by false alarms for other declinations.

To summarize the three methods for measuring the false alarm rate: approximately 20%–33% of spirals and circles have indicated directions that are false alarms. The false alarm rate is 1–5 wedge-shaped sunbeams on flat rock panels for any target declination for one side of one small hill with only one hour of searching. When a site has one real intentional sun dagger, it appears likely that (at the three-out-of-three level) false alarms will appear on other dates.

## 5 WHAT ARE THE CRITICAL AND INTENTIONAL FEATURES OF THE SUN DAGGER?

How can we recognize a sun dagger as being intentional? The best way is to find ethnographic evidence from the direct descendants of the Hohokam and Ancestral Pueblo peoples. This path provides clear answers to some aspects of the sun dagger phenomenon, but little help on other aspects. Another path is to use ordinary astronomy to give clear answers of what is and is not possible. The most general way to identify false alarms is to use the case studies to demonstrate which properties of a sun dagger were required, which were common, and which were not used. From this, each individual proposed sun dagger can be tested for how well they match against this template.

### 5.1 Alignments on the Solstices Are Intentional

The Preston histogram (Figure 1) provides the proof that the solsticial sun daggers are intentional. As in Section 4, not all solsticial sun daggers are intentional, but most of them are. So, if we see a sun dagger with an indicated solar declination of ±23.4° or so, then we have a good expectation that the builders designed the petroglyph as a solar marker.

The cornerstones of the Tohono O'odham and Pueblo calendars are the two solstice dates (Bostwick, 2010; McCluskey, 1977). Further, the archaeology for the Hohokam and Ancestral Pueblo peoples points to the primacy of the solstices (Bostwick, 2010). So the ethnography and archaeology both confirm that intentionally aligned sun daggers would likely point to the Sun at the solstices.

### 5.2 Alignments on the Equinoxes Are Intentional

The Prestons' histogram also proves that the Hohokam and Ancestral Pueblo peoples intentionally constructed petroglyphs to somehow mark the equinoxes. For the numbers above the noise level, only 11% are in the equinoctal bin, with the rest being solsticial. Fountain's statistics show that 32% of his solar markers are equinoctal, which is 40% of the solstice + equinox markers. So we know that the constructed sun daggers marked the equinoxes only 11%–40% of the time.

The significance and importance of the equinoctal timing is already established, but that does not mean that any specific claimed equinoctal sun dagger is intentional. After all, we still have 20%–33% of the sun daggers as unintentional coincidences, just noise caused by the myriad of shadows and petroglyphs. From the Prestons' histogram, for spiral or circular petroglyphs that have an indicated declination on the celestial equator, 33% are false alarms. So each individual claim must be examined on its own merit.

The Picture Rocks Sun Dagger does show a wedge-shaped sunbeam with the apex touching the spiral center on the day of the Equinox. This is a classical intentional sun dagger. Dart's discovery of the man-made chipping of the rock in right-notch demonstrates that the equinoctal sun dagger is intentional. So the Picture Rocks Sun Dagger has an intentional solar marker for the equinoxes.

A resolvable technical problem is to ask how the ancient peoples determined the day of the equinox. Determining the day of the solstice is easy with a horizon calendar and very well attested ethnographically, but there is no easy way to determine the day of equinox. There is no evidence and no plausibility that the equinox was recognized with a count of 91 days after the solstice (e.g., McCluskey, 1977). However, an easy method to base the equinox determination is to use the dates on which the Sun rises to the east. Indigenous cultures throughout the American Southwest have a ubiquitous importance for the cardinal directions and for watching the sunrise positions, so this method is culturally-appropriate and well-attested.

## 5.3 The Wedge-Shape For the Sunbeam Is Important

Is the wedge-shape or a 'dagger-shape' intentional or required for a sun dagger? This question is important for evaluating claims that some specific sun dagger is intentional. For example, if we find a site where a 3-inch diameter circular sunbeam is centered on a spiral on the winter solstice, then should we conclude that this is an intentional sun dagger? Or what if the site has a jagged blob centered on the spiral on the winter solstice? That is, how important is the wedge-shape?

Unfortunately, we do not have ethnographic evidence regarding wedge-shapes or dagger-shapes. This might be expected, because there is no mention of the sun dagger phenomenon at all, so details on the sunbeam shape must also be absent. We know of no wedge-shaped petroglyphs, so this symbol is at least not-common, which weakly suggests that the shape is not important for the pre-contact cultures. In the absence of outside evidence, we can only consider the sun dagger case studies.

For the case studies, we should only examine sites where the center of a spiral has a sunbeam interaction on either the solstice or equinox. The Picture Rocks Sun Dagger has a good wedge shape with opening angles of roughly 20° and 30°. The Fajada Butte Sun Dagger can be characterized as a wedge with an opening angle of around 3°. The Hayden Butte spiral petroglyph has a 40° wedge on the summer solstice (Bostwick, 2010: Figure 17). The South Mountains spiral has a wedge of roughly 25° opening angle on the summer solstice (Bostwick, 2010: Figure 18). The spiral at Site 27 of Preston and Preston (2005: Figures 1 and 6), in north-central Arizona, has a wedge with opening angle of around 12° that touches the center on the summer solstice. In summary, these case studies have a simple wedge-shape of light with an opening angle of 3°–40°.

A related question is whether a straight shadow line (i.e., a wedge opening angle we can identify as 180°) passing across the spiral was ever an intentional solar marker (e.g., Figure 5 in Preston and Preston, 2005). Well, it certainly is not wedge-shaped or dagger-shaped. But is it an intentional solar marker? The answer really has to be "no". The problem is that the Sun's motion across the sky will usually make a shadow-line pass completely across the spiral, so there is no coincidence and nothing special in having the line pass over the spiral center. Further, the general case of the Sun 'rising' over a rock panel will frequently have a nearby rock casting a shadow-line passing over the spiral, resulting in a very high false alarm rate. Critically, for testing the sun dagger hypothesis, these shadow-lines cannot mark either solstices or equinoxes, because the shadow lines must always have similar passage across the spiral for dates weeks and months before and after the solstice/equinox. That is, a simple shadow-line cannot be used as a solar marker.

So some sort of a wedge-shape appears to be a requirement for intentional sun daggers. For all cases, the apex has an opening angle of ≤40°. Straight lines cannot be used as solar markers, while triangles with opening angles >40° have not been used in any of the confidently recognized sun daggers.

## 5.4 It Is Important To Interact With the Center of the Petroglyph Design

The basic pattern for the sun daggers is to have the sunbeam interact with the center of the spiral. This makes for easy construction in ancient times and easy interpretation by the users. The use of a spiral (or circle) center to indicate the importance of a singular point of importance has obvious symbolic merit, and we expect that this point is universal. However, we have no ethnographic evidence pointing to whether the sunbeam must interact with the center for an intentional sun dagger.

Without outside evidence, we can only look to the case studies of individual sun daggers. For this, we have to restrict the analysis to cases that are confidently known to be intentional sun daggers independently from the position of the sunbeam with respect to the spiral (or other petroglyph symbol). With this, we collected reports for which wedge-shaped sunbeams interact with spirals or circles on either the solstices or equinoxes. Within this collection, there might be a small fraction of random coincidences, but we can get a good idea as to the fraction of sunbeams that interact with the center. We take a center-interaction to be when the Sun illuminates the center of the spiral or circle with any part of the wedge.

Unfortunately, we have found only five sun dagger sites with adequate published information. The Picture Rocks Sun Dagger has the apex of the wedge of light on top of the spiral center on the equinoxes and it has the right edge of the wedge touching the spiral center on the summer solstice. The Fajada Butte Sun Dagger has the center of the very narrow wedge of light touching the center of the spiral on the summer solstice, while a different spiral interacts with a very narrow wedge of light touching the center on the equinoxes. The summer solstice wedge at Hayden Butte is touching inside the innermost coil (Bostwick, 2010). The South Mountains spiral has a wedge of light covering the center on the summer solstice (*ibid.*). Site 27 of Preston and Preston (2005) has the base of a thin wedge covering the spiral center on the summer solstice. Although it is small number statistics, six out of six have the interaction with the center.

So it appears that a sunbeam/center interaction is required for an intentional sun dagger. But this is not a strong conclusion, because we only have six cases. More detailed case studies are needed.

## 5.5 Spirals and Circles Are Common, But Other Symbols Are Used

Many sun daggers are constructed with a petroglyph of a spiral. The popular expectation and statement is that spirals are somehow representative of the Sun, thus the symbol would function as an intentional connection from the petroglyph to sunbeams. Unfortunately, the dominant meaning of a spiral in the ethnographic record for Pueblo peoples is as a whirling wind, whirling flood waters, or the migration of peoples (Zeilik, 1985a). However, in just one case, a spiral petroglyph is said to represent the Sun, inside a *kiva* at the Jemez pueblo (*ibid*.). Given the possibility that spirals might have multiple meanings and those meanings might change over time, the disconnect between spiral and Sun is not decisive.

The Picture Rocks Sun Dagger and the Fajada Butte Sun Dagger both have their sunbeams interacting with a spiral shaped petroglyph. Fountain and the Prestons show pictures of six other interactions with spirals. For the Hohokam, Bostwick (2010) shows pictures of three spirals interacting with light wedges. So spirals in sun daggers are both a common case as well as the popular public paradigm.

But the original case, as told to A. Sofaer at a conference just before her discovery on Fajada Butte, was of a sun dagger in Baja California that had the light wedge interacting with a stick figure with horns on the head, perhaps an image of a shaman, with no spirals anywhere near (Krupp, 2000). And the winter solstice marker at Paint Rock, in central Texas, shows a narrow and simple wedge of light with its apex at the center of what looks like a heraldic shield of post-contact origin (Houston and Simonia, 2015). So apparently, a spiral petroglyph is not required for an operational sun dagger.

Preston and Preston (2005) include both spirals and circles in their statistics. So the proof of intentionality includes both circles and spirals. Fountain (2005) gives statistics for this sample of solar markers, with circles and spirals constituting 37% of his sample, anthropomorphic figures constituting 21%, zoomorphic figures providing 15%, and the remaining 27% are described as ‘miscellaneous’. So these two surveys already show that spirals, circles and various other figure were used for intentional sun daggers, although spirals are most common.

The answer to these issues can be improved by many more specific case studies, after which detailed statistics of many properties can be used to identify significant patterns that go with demonstration of intent. In the meantime, some of the questions can be answered with varying confidence. The widespread and common appearance of the basic sun dagger characteristics (wedge-shaped lights on spirals at equinox/solstice times only around the American Southwest) already gives good confidence for intentionality, whether or not variations on that pattern are included. Further, the similarity of interactions with non-spiral petroglyphs with the proven-intentional sunbeam/spiral interactions, plus the high frequency of use for non-spiral petroglyphs, is already an adequate demonstration of intentionality. So a number of petroglyph patterns have been used for intentional sun daggers, although the spiral pattern is the most common.

## 5.6 The Time of Day Is Not Important

The Paint Rock Sun Dagger in central Texas has the sunbeam wedge crossing the center of the emblem almost exactly at the time of local noon (Houston and Simonia, 2015). On this basis, Houston and Simonia elevated the coincidence with noon as being one of the highest criterion for determining intent. But is this reasonable?

Perhaps the original basis for the idea of the importance of 'noon' is the statement in Sofaer et al. (1982) that the most prominent interaction occurred at mid-day. But as Zeilik (1985a: S74) points out:

> Unfortunately, the public media have picked up on this diagram [in Sofaer et al.] and converted the phrase "at midday" into "noon".

In a related claim, Sofaer and Sinclair (1987) note that five out of twenty petroglyphs on Fajada Butte have various sorts of interactions with light beams all within twenty minutes of local noon, and they took the improbability of this to imply that the Ancestral Puebloans were intentionally marking local noon. No motivation or precedent or ethnography was advanced to support the probability argument. McCluskey (1988) proved that the rate of noon-time interactions is actually the same as expected by random chance. So the Fajada Butte petroglyphs are not evidence for the importance of noon-time events.

A strong reason to doubt the importance of local solar noon is that no ethnography points to any recognition or use of local solar noon, or anything like it, by any group in the American Southwest. (Whereas, many sources point to intense observations of the low-altitude Sun for a variety of applications.) However, one referee points to a Fred Katobie mural from 1933 inside the Desert View Watchtower overlooking the Grand Canyon, where the small pair of chairs at the top of the rainbow-like arc indicating the Sun's daily path can be considered as a recognition of solar noon. But this isolated recognition of noon is ambiguous and modern. With the only recognition of solar noon being weak and modern, we have a strong case that the old cultures of the American Southwest gave little importance to the concept of solar noon, and hence would be very unlikely to commemorate this in any sun dagger.

With our case study of the Picture Rocks Sun Dagger, the best summer solstice wedge is at a time near 11:47 a.m., while the time of local noon is 12:26 p.m. This solsticial sun dagger is confidently one of the intentional cases, and this happens 39 minutes before noon. The equinoctal sun dagger is also confidently intentional, and it is 68 minutes before local noon. The Fajada Butte Sun Dagger has the solsticial wedge crossing the center of the spiral only for a few minutes around 11:11 a.m. local solar time, which is to say that the proto-type sun dagger only works for a time 49 minutes before noon. The Jackrabbit Sun Dagger in northern Arizona has the wedge starting to touch the spiral 3 hours and 54 minutes after local solar noon on the summer solstice (Bates and Coffman, 2000). The equinoctal sun dagger in Rarick Canyon has the point crossing the center of the spiral 30 minutes before noon (ibid.). The original sun dagger, near La Rumorosa in Baja California, has the interaction at 40 minutes after sunrise on the solstice (Williamson, 2015a). With a large database, Fountain (2005) concludes that "Such indirect solar markers may receive interactions at any time from sunrise to sunset." With this, we see that the there is nothing special about noon-time, or any other time.

Nevertheless, there is likely to be somewhat of a bias on the builders' part for avoiding the afternoon, at least for the summer solstice and autumnal equinox sun daggers. The reason is simply that a common weather pattern throughout the American Southwest is to have clouds building up to scattered thunderstorms, starting in the afternoon, commencing around the time of the summer solstice and ending around the September equinox.

## 5.7 Cross-Quarter Days Are a Modern Eurocentric Fantasy

Various claims have been made that sun daggers mark the cross-quarter days (e.g., Fountain, 2005). Cross-quarter days are also called mid-quarter days or Scottish quarter days. These are dates halfway between the solstices and equinoxes; 4 February, 6 May, 6 August, and 5 November. With these added to a calendar, the year is divided into eight equal time intervals. On these dates, the Sun's declination is close to ±16°. So the question is whether sites that point to $\delta_{\odot}$ near ±16° are intentional sun daggers?

One quick and sure refutation is that the Prestons' histogram (see Figure 1) does not have peaks at ±16°. This shows that the various claims for a cross-quarter day alignment are just random chance, the expected noise from many shadows and petroglyphs on rock faces throughout the American Southwest.

Historically, the concept and usage of cross-quarter days is from the Celtic calendar, and is entirely confined to the British Isles and descendants (Ruggles, 2005). The trouble started when early archaeoastronomers looked for alignments in the British Isles relating to cross-quarter days, inevitably found alignments (fully consistent with random noise), and their publications made for archaeoastronomers worldwide including cross-quarter days in their toolkits. With cross-quarter days in their toolkits, researchers on sun daggers will inevitably investigate the possibility of alignments on those days, and will inevitably find apparent alignments for those days (or any other day) from just random chance coincidences. When researchers started considering the archaeoastronomy possibilities in the New World, they had a knee-jerk check for cross-quarter days, and inevitably found 'exciting results'. Thus, a European calendar trait was inappropriately transmitted to the American Southwest in modern times. It is not wise or reasonable to search for a Celtic calendar in pre-contact Arizona.

More in particular, we know nearly full details of the Hohokam and Ancestral Puebloan calendars and the Pueblo and Tohono O'odham calendars (Bostwick, 2010; McCluskey 1977; Zeilik 1985a), and there is no indication of anything like cross-quarter days, nor any importance attached to the four dates. With this, the existence of cross-quarter day alignments is certainly *not intentional*. That is, cross-quarter days for sun daggers are just a modern Eurocentric fantasy.

## 5.8 Lunar Alignments Are a Useless Modern Fantasy

'Theorists' have proposed that sun daggers were intentionally used by the old cultures to observe lunar standstills (also called 'lunastices'), where the Moon is at its extreme declinations of ±28.6°. The idea is presumably that the observers would notice the one or two nights every 18.6 years when the Moon was at these extreme declinations and cast an observable moon dagger at some part of the rock face that was marked in some assertedly special manner. There is zero utility for any person or society that can be derived from anything related to the lunar standstills. There is zero evidence for this idea that lunar standstills were marked by sun daggers (or anything else), other than the supposed existence of the alignments, but that is a circular proof of assuming that which is being tested.

The Picture Rocks Sun Dagger could never have been used as a moon dagger for the Moon at any declination. This is proved by the utter inability for the human eye to see either the moonbeam with its edges, or the spiral petroglyph, even with a Full Moon.

What about other sun daggers? Most other sun dagger sites have better conditions for seeing a moon dagger. In particular, the observer can usually get much closer to the panel, and the illumination onto the panel is usually not oblique. The determination of the visibility of each proposed moon dagger will require carefully-made special-purpose observations around Full Moon for each site. We know of only one other site with real observation of a moon dagger, and that is for one night on Fajada Butte, where the moonbeam is reported to be easily visible (Sofaer et al., 1979). Critically, this report did not say whether the spiral was visible for a human standing in the entrance to the alcove, and we are dubious as to whether the petroglyph could be seen because the spiral is in a completely dark alcove except for a narrow sliver of faint moonlight. (The published photograph is irrelevant for this because long-exposure photography will record greatly better than the human eye.) So advocates of lunar standstills must provide observational evidence that a moon dagger and petroglyph is actually visible, all on a case-by-case basis.

A simple demonstration that the ancient desert dwellers did *not* construct intentional moon daggers for the lunar standstills is the lack of any peaks at ±28.6° in the Prestons' histogram (see Figure 1). So whether or not it is even *possible* to use moon daggers, they were *not* so used.

Further, the ethnographic evidence strongly shows that the people throughout the Southwest had zero knowledge of or interest in any phenomena even remotely likened to lunar standstills (Bostwick, 2010; Carlson, 1987; Williamson, 2015b; Zeilik, 1985a, 1989). This strong conclusion can be generalized to the entire world, for all times before 1912. (In 1912, the entire concept of lunar standstills was invented by Boyle Somerville, as he desired to find astronomical alignments amongst the many standing stones at the Scottish site of Callanish; see Ruggles, 2005.) For both the American Southwest and for the entire world, many of the top workers in the field have combed massive numbers of written and ethnographic records, seeking anything that has any pre-1912 peoples being aware of or interested in any lunar phenomena that is even vaguely connected to the lunar standstill idea (Schaefer, 2017a). Nothing was found. The conclusion is that no known culture or person anywhere, for any time before 1912, had knowledge of or interest in lunar standstills. Thus, we can be very confident that the Hohokam and Ancestral Puebloans had no knowledge of or interest in lunar standstills.

### 5.9 Summary of Intentional Features of Sun Daggers

The establishment of *intentionality* is critical for evaluating individual sun dagger sites, especially because we know that the false alarm rate is significant. Any sun dagger that was not made with the intention of the builders is useless and uninteresting. So we need to somehow get into the minds of the petroglyph carvers. The Preston histogram proves that most of the solsticial and equinoctal sun daggers are intentional. But there is a large false alarm rate, so we need to evaluate each sun dagger case individually. We do not have ethnographic reports for individual sun daggers, so we can only go by the properties of each individual sun dagger. For this, we have evaluated many properties for whether they are required, common, possible, or impossible for an intentional sun dagger. With this profile, we can evaluate each individual site for the likelihood that the builders were making a solar marker.

Houston and Simonia (2015) offer a much needed ray of hope by presenting a quantitative scale, which they called the 'Solar Marker Matrix of Intentionality'. This consists of adding points (each on a 1–5 scale) for four questions: "What day of the year is marked?", "'What time of day is used?", "What type of interactions are seen?", and "What supporting evidence is available?" This is a reasonable approach. However, to establish the scales for assigning the 1–5 points, they had to establish which features of the sun daggers are intentional. Unfortunately, many items on their scale need revision. For example, the evidence indicates that there was no intentionality concerning the time of day, so that a quarter of the total score given by the Matrix needs to be withdrawn. And the cross-quarter days need to be eliminated from contributing to the score, or perhaps to be given negative points.

Let us summarize what we have learned for the properties of an intentional sun dagger. Most important, the intentional sun daggers all are indicating solar declinations of ±23.4° or 0°, to within roughly 2° accuracy. Sun daggers pointing at any other declinations are false alarms. With less strong confidence, intentional sun daggers use wedge-shaped sunbeams with opening angles of ≤40° where the wedge touches the center of the petroglyph. The petroglyph of intentional sun daggers are most commonly spirals, but circles and other figures have been used. The time of day is irrelevant.

## 6 WHAT WAS THE PURPOSE AND USAGE OF THE SUN DAGGERS?

We can imagine many possible purposes for the sun daggers. A wide variety of claimed and implied uses have been published. Here, we will address various possible purposes in light of the evidence from the Picture Rocks Sun Dagger, other sun daggers, archaeological evidence, ethnographic reports and astronomical possibilities.

### 6.1 Calendars

The most popular idea is that the sun daggers were used as time checks for setting the solar calendar (e.g., Magli, 2009; Sofaer et al., 1979). We can easily imagine some elderly 'priest' visiting a sun dagger daily, finally coming back to announce the start of the year, when to plant crops, and the dates of up-coming celebrations. But this vision is certainly wrong, for two reasons:

(1) The sun daggers have very poor calendrical accuracy, so cannot be used for any practical setting of calendars. Ethnographically, we know that the solstices are the critical dates in the solar calendar for rituals and celebrations in the Southwest (Bostwick, 2010; Zeilik, 1985c), and this is confirmed by the peaks in the Prestons' histogram. But sun dagger phenomena are only sensitive to the changing solar declination, and the Sun's declination changes only very slowly around the solstices. In particular, the Sun's declination is within 0.5° of its extreme for 27 days centered on the solstices, is within 0.2° of the extreme for 17 days, and is within 0.1° of the extreme for 11 days. Our data for the Picture Rocks Sun Dagger show no significant variation in the sunbeam placements over a 48 day interval centered on the summer solstices. Further, even for the equinoxes, with the Sun moving the fastest in declination, our practical observations show that the date of equinox cannot be determined to better than a week or so. Similarly, for the Fajada Butte Sun Dagger, Zeilik (1985a) concludes: "To use the site to anticipate the summer solstice to its actual day requires a precision that it does not and cannot display." Similar results are known for four sun dagger sites in California, where the light-and-shadow effects seen at solstice are unchanged for intervals of 16 – 32 days (Krupp, 1994). In all, we confidently know that sun daggers cannot be used in practice to set calendars to an accuracy of a week or so (Krupp, 2015b).

(2) Ethnographically, we already know that the Hohokam and the Ancestral Puebloans used sunrise calendars (watching the position of sunrise on the eastern horizon). With a known place to stand and a distant horizon with many topographic features, the date of solstice can be easily determined with an accuracy of one day, and the cultural practice was to *anticipate* the upcoming solstice dates with good accuracy for ritual preparations. So we already know the method by which the solar calendar was regulated. The cultures of the old Southwest would not have used sun daggers as calendrical tools when we already know that they actually used other tools of much greater accuracy and relevance.

## 6.2 Astronomical Tools

Another evocative idea is that the sun daggers might have been used for technical observations that we could characterize as astronomical science (e.g., Fazier, 1979). For example, Widner (2016: 155) talks about an "Anasazi Einstein". We can imagine some ancient 'astronomer' repeatedly visiting the sun dagger to collect data on the length of the year, on the obliquity of the Earth, and on the Moon's nodal cycle. But this vision is certainly a fantasy, for two reasons:

(1) A sun dagger *cannot* be used for any non-trivial astronomical purpose.

(2) Ethnographically, we know that the peoples of the American Southwest had no knowledge of or interest in any issues like the length of the year in days, the obliquity of the ecliptic or the Moon's nodal cycles. The next three paragraphs in this section will detail the reality that sun daggers were not astronomical tools for a variety of conceivable astronomical questions.

Could the sun daggers have been used to provide a tool to measure the number of days in a solar year? The idea would be that a Sun Priest notes a specific date with the sun dagger, then counts the 365 or 366 days until the same light/shadow configuration occurs, thus measuring the number of days in a solar year. The trouble is that any given solar day, even at the equinoxes, can be pinned down to only ±4 days, so the year-length would be measured with an accuracy of ±4*√2 or near ±6 days. (In principle, keeping a count over half-a-century would allow an accuracy of better than one day, but such a modern solution is an anachronism and is not culturally appropriate for the Hohokam or Ancestral Puebloans.) If an ancient 'astronomer' desired to measure the year length, they would just simply have used their extant horizon calendar, with much greater accuracy. The ethnography of the descendants of the Hohokam and Ancestral Puebloans contains no knowledge of day counts for a year (Lopez, 2020; Stirling, 1945), and no practice or ability to count to large numbers (McCluskey, 1977; Russell, 1908; Zeilik, 1985a). In all, the sun daggers could not be used and were not used to count the days in a solar year.

Could the sun daggers have been used as an astronomical tool to measure the obliquity of the ecliptic? The idea would be that a Sun Priest would note the positions of solsticial and equinoctal sunbeams so as to calculate the extreme declinations of the Sun and to realize the tilt of the Earth's orbit. The trouble is that the rock surfaces are irregular, complex, and ill-defined so that even a modern geometer would have difficulty working from the sunbeam positions to calculate the angle. If an ancient 'astronomer' desired to measure the obliquity with a device, then the design certainly would have been for something greatly simpler. In any case, the whole idea that the Southwestern peoples were trying to measure anything even vaguely like the obliquity is just an anachronism. Ethnographically, the Pueblo peoples and the Tohono O'odham had no concept or usage of anything related to obliquity, orbits, angles or noon-time zenith distances (e.g., Bostwick, 2010; Cushing, 1967; Lopez, 2020; McCluskey, 1977; Russell, 1908; Stirling, 1945; Zeilik, 1985a). Nor did they have any use or practice of accurate measures of distances or angles. So, the sun daggers could not be used and were not used to measure anything like the obliquity.

Could the sun daggers have been used as an astronomical tool to measure the Moon's nodal cycles? The idea would be that a Sun Priest would measure the nodal period, or the inclination of the Moon's orbit, or the amplitude of lunar nutation. A problem for the Picture Rocks Sun Dagger is that the moon dagger phenomenon is invisible, and we expect this to be a common trouble. Lunar nutation causes a complex wobble with an amplitude 0.0025°, greatly too small for detection with a moon dagger. Further, our detailed simulation of lunar visibility shows that the time intervals be-tween observed northern extremes varies from 17.7 to 21.0 years, with a lunar declination range of 1.7°. Further, lunar declination extreme moon daggers would only be visible on one night around Full Moon, in the one month close to the winter solstice (for a northern lunar standstill), in only one year out of every ~19 years, and that night has about a 40% chance of being cloudy in the few minutes when the moonbeam passes over the petroglyph. In the absence of a theoretical framework, written records, and any motivations to record such observations over centuries, the discovery and recognition of any such phenomena is impossible. Ethnographically, there is zero knowledge or practice of anything like lunar stand-stills or lunar nodal cycles anywhere in the American Southwest (Bostwick, 2010; Carlson, 1987; Ruggles, 2005; Williamson, 2015b; Zeilik, 1985a). In all, sun daggers could not be used and were not used to measure or recognize lunar nodal cycles because such are impossible anachronisms.

## 6.3 Public Ceremony

Another idea is that perhaps the sun daggers were used as part of some public ceremony/ ritual/celebration. We can easily imagine the local population coming together from the countryside and village, gathering around a sun dagger, presided over by a local 'priest', with a joyous celebration kicked off when the cheering crowds saw the sunbeam touching the center of the spiral. But this vision is just a modern fantasy because:

(1) The sun daggers are poor public spectacle. The Picture Rocks Sun Dagger lasts under five minutes, the petroglyph is hard to see in full sunlight on a rock panel high above the wash, and the sunbeam wedge is just a fraction of a confusing and complex large array of sunbeams. The Fajada Butte Sun Dagger is high atop a nearly inaccessible set of mostly vertical cliffs, and can be viewed for only four minutes from a ledge by peering into a small slot behind the rock slabs.
(2) There is zero archaeological or ethnographic support for public viewing of anything in the sky.
(3) In particular, the observations of the Sun near the horizon for calendrical purposes was always done by lone Sun-watchers, as known from many sources for the Puebloans (Zeilik, 1985b) and the Tohono O'odham (Bostwick, 2010).

## 6.4 Sun Shrines

The fourth and last of the proposed purposes for intentional sun daggers is that the petroglyph and sunbeam phenomena were a small part of a Sun Shrine, wherein the site was used by a local Sun Priest for small and private rituals in devotion to the deities, as pleas for good harvests and general wellbeing. The use and character of these Sun Shrines have many descriptions from the ethnographies of the latter-day descendants.

Fewkes (1892) gives a first-hand account of a Sun Priest from Hano visiting a Sun Shrine for the summer solstice. In this case, the visit started before dawn, when a line of prayer feathers was planted in the back of the shrine. The priest sat facing south, awaiting the rising of the Sun, offering a short prayer to bring good luck to his people. When the Sun appeared, the priest cast cornmeal onto the feathers and to the east. The private ritual was similar for the winter solstice, although at a separate Sun Shrine. The rituals and physical settings differ from village to village, but the character remains the same. Zeilik (1985b) characterizes Sun Shrines as being places where offerings to the Sun are deposited, located far away from villages, and with natural or manmade piles of rocks, sometimes with rock art, and are used for commemoration of key times in the ritual or planting calendar.

These Sun Shrines are distinct from Sun-watching stations (Zeilik, 1985b). Sun-watching stations are for practical observing of the Sun for calendrical purposes, and are usually near to the living areas, and for which offerings to the gods are not deposited. For the Puebloans, early and strong ethnography from several sources (e.g., Cushing, 1967; Ellis, 1975) state that the Sun-watching stations are indicated by marking the site with petroglyphs (or pictographs) depicting the crescent Moon, a star, the Sun's disk, and a hand, with this unique assemblage being found throughout the Southwest. It is certainly possible that Sun-watching stations were indicated by other means.

The most famous Sun-watching station is the one at Peñasco Blanco, in Chaco Canyon, as marked by pictographs showing the crescent Moon, a star, the Sun's disk, and a hand. This is the infamous 'Crab Supernova Pictograph', now widely highlighted in astronomy textbooks and across the internet. The only positive evidence was that the Crab Supernova appeared next to a crescent Moon on 5 July 1054 and the Peñasco Blanco site was occupied from approximately AD 900 to 1125. But more recent evaluations show the specific pictographs to likely postdate the Chaco occupation, and hence to not be any commem-oration of the supernova (Krupp, 2015a). Further, Krupp (2014) tracked down the other cases of star-plus-crescent rock art claimed to be the Crab, and demonstrated that they are not 'supernova', but rather various other known symbols. For the specific case of Peñasco Blanco, the sure refutation of the supernova hypothesis is that the star-plus-crescent is actually a tight configuration of glyphs displaying the crescent Moon, a star, the Sun's disk, and a hand, so the rock art certainly indicated a Sun-watching station (Ellis, 1975). To see the eastern horizon, the ancient observer would had to have moved several yards away from the cliff base, up a nearby berm to about the same height as the high-up pictographs. In all, we can be very confident that this most famous case of rock art is not of a supernova, but rather is an ordinary Sun-watching station.

The Sun Shrine explanation for the purpose of sun daggers is a good fit. They are low precision devices for commemorating the solstice Sun, widely scattered throughout the American Southwest, but not close to any villages. Sun daggers and Sun Shrines are on rock cliffs and rock piles with apparent solar symbols as rock art. As the exact opposite of the first three proposed purposes (see Sections 6.1 to 6.3), the Sun Shrine explanation has ample and detailed ethnographic reports telling us explicitly what is going on.

The experienced leaders of our field (e.g., Krupp, 1983; Ruggles, 2005; Williamson, 2015a; Zeilik, 1985b, 1985c) have all concluded that the sun daggers have their purpose serving as one component of a Sun Shrine. So we have a confident answer for knowing that the sun daggers served as part of a private commemoration site, used by local Sun Priests or Sun Watchers to make offerings and prayers to the Sun at the critical times of the solar calendar.

## 7 CONCLUSIONS

We present a full case study of the Picture Rocks Sun Dagger. This is only the second full case study in the literature, with the first being 41 years ago for the Fajada Butte Sun Dagger. We use our case study, plus partial information available in the literature, to evaluate sun daggers and their properties as products of *intention* by the petroglyph-makers.

We found that:

(1) The Picture Rocks Sun Dagger operates for two solar declinations (+23.4° and 0°), both of which are certainly intentional. The summer solstice event has a large sunbeam wedge with an apex opening angle of ~20° that barely touches the spiral center. The equinoctal event has a sunbeam with a wedge-shape with an opening angle of ~30° whose apex just touches the center of the spiral at the end.

(2) These two sun dagger events were intentionally made by the Hohokam. They started by recognizing a wedge-shaped sunbeam on their targeted hill on the summer solstice, decided on a center, and pecked the spiral. This all could have been done in one day by one ordinary person. Later, they chipped the rock edge to create the right-notch and the equinoctal sun dagger. We know these were intentional creations because the Prestons' histogram proves that most solsticial and equinoctal sun daggers are intentional, because the Hohokam chipped the rock to create the sun dagger wedge-shaped sunbeam on the equinox, and because both sun daggers share the critical properties (a wedge-shaped sunbeam interacting at the center of a spiral petroglyph) with many other equinoctal/solsticial sun daggers.

(3) The false alarm rate is 20%–33% for spirals or other petroglyph symbols. For a single side of a small hill for one hour of searching, the rate of finding a wedge-shaped sunbeam projected onto a flat rock face was 1–5. This makes the task of finding a position to create a sun dagger easy for anyone. It also means that a substantial fraction of claimed and published sun daggers are not intentional and hence of no interest to anyone.

(4) A critical feature for intentional sun daggers is that they point to target solar declinations of –23.4°, 0.0°, and +23.4°, as proven from the Prestons' histogram. The cross-quarter days and the lunar standstills are not indicated, nor any dates/declinations other than the solstices and equinoxes.

(5) For these intentional sun daggers, all consist of sunbeams with a simple wedge-shape, with apex opening angle of 40° or smaller, that covers the center of the symbol. The majority of the intentional sun daggers have the petroglyph being a spiral or a circle. However other symbols have been used with intention, including a head of a stick figure and some sort of a heraldic crest. The time of day is not relevant.

(6) The original intended purpose of the sun dagger-makers was not for any setting of calendars, was not for anything we would call as astronomy or science, and was not used for any public ceremony. Rather, ethnographic and archaeological evidence suggests sun dagger sites were used mostly as Sun Shrines for private ceremonies making offerings and prayers to the gods, usually for a good harvest.

## 8 ACKNOWLEDGEMENTS

We thank Allen Dart for discussions, his original site report, and his private communication of his 2009 discovery that the 'right-notch' of the Picture Rocks Sun Dagger had one of its edges chipped away by human hands. We thank Martha Schaefer for high resolution still pictures. Rolf Sinclair, Kirk Astroth and Allen Dart all provided detailed corrections, additions, commentary, and encouragement for a draft of this paper.

## 9 REFERENCES

AZSITE, 2020. *Arizona's Cultural Resource Inventory*. Arizona State Museum (http://azsite3.asurite.ad.asu.edu/Azsite/about.html).

Bates, B.C., and Coffman, L., 2000. Potential astronomical calendars and a cultural interpretation thereof along the Palat'kwapi Trail of North Central Arizona. In Esteban and Belmonte, 125–131.

Blitz, J.H., 1988. Adoption of the bow in prehistoric North America. *North American Archaeologist*, 9, 123–145.

Bostwick, T.W., 2010. Exploring the frontiers of Hohokam astronomy: tracking seasons and orienting ritual space in the Sonoran Desert. *Archaeoastronomy*, 23, 165–189.

Carlson, J.B., 1983. The selling of Fajada Butte: an anacalypsis. *Archaeoastronomy*, 6, 156–160.

Carlson, J.B., 1987. Romancing the stone, or moonshine on the sun dagger. In Carlson and Judge, 71–88.

Carlson, J.B., and Judge, W.J. (eds), 1987. *Astronomy and Ceremony in the Prehistoric Southwest.* Albuquerque, Papers of the Maxwell Museum of Anthropology, No. 2.

Cushing, F.H., 1967. *My Adventures at Zuni,* Palmer Lake (Colorado), Filter Press.

Dart, A., 2009. Description of Picture Rocks Petroglyph Site, AZ AA:12:62(ASM). Site report available at https://www.google.com/url?sa=t&rct=j&q=&esrc=s&source=web&cd=&ved=2ahUKEwjL-div99npAhWiPH0KHWrhCagQFjAAegQIAxAB&url=http%3A%2F%2Fdesertrenewal.org%2Fwp-content%2Fuploads%2F2018%2F05%2F00-Picture-Rocks-site-description-20090215.pdf&usg=AOvVaw1aJzj4TjE3kPen0WIBQUB2

Ellis, F.H., 1975. A thousand years of the Pueblo Sun-Moon-star calendar. In Aveni, A.F. (ed.). *Archaeoastronomy in Pre-Columbian America*. Austin, University of Texas Press. Pp. 59–87.

Esteban, C., and Belmonte, J.A. (eds.), 2000. *Oxford VI and SEAC 99, Astronomy and Cultural Diversity*. Laguna (Canary Islands), Museo de la Ciencia y el Cosmos.

Fewkes, J.W., 1892. A few summer ceremonials at Tusayan Pueblos. *Journal of American Ethnology and Archaeology,* 2, 1–160.

Fountain, J., 2005. A database of rock art solar markers. In Fountain and Sinclair, 101–108.

Fountain, J.W., and Sinclair, R.M. (eds.), 2005. *Current Studies in Archaeoastronomy, Conversations Across Time and Space; Selected Papers from the Fifth Oxford International Conference at Santa Fe, 1996*. Durham, Carolina Academic Press.

Frazier, K., 1979. The Anasazi sun dagger. *Science 80*, 1, 56–67.

Houston, G.L., and Simonia, I., 2015. Pictographs at Paint Rock, Texas: exploring the horizon astronomy and cultural intent. *Journal of Astronomical History and Heritage*, 19, 3–17.

Krupp, E.C., 1983. *Echoes of the Ancient Skies: The Astronomy of Lost Civilizations*. Oxford, Oxford University Press.

Krupp, E.C., 1994. Archaeoastronomy unplugged: eliminating the fuzz tone from rock art astronomy. In Hamann, D., et al. (eds.). *International Rock Art Congress 1994, Volume 3*. Phoenix, American Rock Art Research Association (*American Indian Rock Art, Volume 21*). Pp. 353–369.

Krupp, E.C., 2000. Rock art and astronomy in Baja California. In Esteban and Belmonte, 133–139.

Krupp, E.C., 2014. Star trek: the search for the first alleged Crab supernova rock art. *Bulletin of the American Astronomical Society*, meeting 223, 437.01.

Krupp, E.C., 2015a. Crab Supernova rock art: a comprehensive, critical, and definitive review. *Journal of Skyscape Archaeology*, 1, 167–197.

Krupp, E.C., 2015b. Rock art of the Greater Southwest. In Ruggles, 593–606.

Lopez, C., 2020. Summer solstice notes (https://camillussite.wordpress.com/summer-solstice/).

Magli, G., 2009. *Mysteries and Discoveries of Archaeoastronomy*. New York, Copernicus Books & Springer.

McCluskey, S.C., 1977. The astronomy of the Hopi Indians. *Journal for the History of Astronomy*, 8, 174–195.

McCluskey, S.C., 1988. The probability of noontime shadows at three petroglyph sites on Fajada Butte. *Archaeoastronomy (Journal for the History of Astronomy Supplement)*, 19(12), S69–S71.

Newman, E.B., Mark, R.K., and Vivian, R.G., 1982. Anasazi solar marker: the use of a natural rockfall. *Science*, 217, 1036–1038.

Preston, R.A., and Preston, A.L., 2005. Consistent forms of solstice sunlight interactions with petroglyphs throughout the prehistoric American Southwest. In Fountain and Sinclair, 109–120.

Reed, P.F., and Geib, P.R., 2013. Sedentism, social change, warfare, and the bow in the ancient Pueblo Southwest. *Evolutionary Anthropology*, 22, 103–110.

Ruggles, C., 2005. *Ancient Astronomy: An Encyclopedia of Cosmologies and Myth*. Washington, Government Printing Office.

Ruggles, C.L.N. (ed.), 2015. *Handbook of Archaeoastronomy and Ethnoastronomy.* New York, Springer.

Russell, F., 1908. *The Pima Indians*. Santa Barbara California, ABC Clio.

Schaafsma, P., 1979. Personal communication reported in Sofaer et al.,1979.

Schaefer, B.E., 2017a. The utter failure of the lunar standstill myth in archaeoastronomy. Paper presented at the joint 10th Meeting of the Inspiration of Astronomical Phenomena (INSAP X), 11th Oxford Symposium of Archaeoastronomy and Ethnoastronomy (Oxford XI) and 25th Annual Meeting of the European Society for Astronomy in Culture (SEAC XXV), Santiago de Compostela, Spain, 18th–22nd September.

Schaefer, B.E., 2017b. *The Remarkable Science of Ancient Astronomy*. 24-lecture course in many formats, see https://www.thegreatcourses.com/. The Great Courses, Chantilly, Virginia. Lectures 5 and 23.

Schaefer, B.E., 2018. The early astronomy toolkit was universal. *Bulletin of the American Astronomical Society*, meeting 231, 108.01.

Schaefer, M.W., Schaefer, B.E., and Stamm, J., 2020. Astronomy Picture of the Day. 5 August 2020 (https://apod.nasa.gov/apod/ap200805.html).

Sofaer, A., and Sinclair, R.M., 1987. Astronomical markings at three sites on Fajada Butte. In Carlson and Judge, 43–70.

Sofaer, A., Sinclair, R., and Doggett, L.E., 1982. Lunar markings on Fajada Butte, Chaco Canyon, New Mexico. In Aveni, A.F. (ed.), *Archaeoastronomy in the New World*. Cambridge, Cambridge University Press. Pp. 169–181.

Solstice Project, 1982. *The Sun Dagger, The Story of America's Stonehenge*. Narrator Robert Redford (https://www.google.com/search?client=firefox-b-1d&q=fajada+butte+sun+dagger+documentary+robert+redford).

Stirling, M.W., 1945. Concepts of the Sun among American Indians. Washington, Smithsonian Institution Annual Report. Pp. 1–12.

Widner, K., 2016. *The Anasazi of Chaco Canyon: The Greatest True Mystery of the American Southwest*. Boulder City (Nevada), Shadowplay Communications.

Williamson, R., 2015a. Sun-dagger sites. In Ruggles, 621–628.

Williamson, R., 2015b. Pueblo ethnoastronomy. In Ruggles, 641–648.

Winters, R., 2004. *The Green Desert: A Silent Retreat*. New York, Crossroads.

Wright, A.M., 2014. *Religion on the Rocks: Hohokam Rock Art, Ritual Practice, and Social Transformation*. Salt Lake City, University of Utah Press.

Zeilik, M., 1985a. A reassessment of the Fajada Butte solar marker. *Archaeoastronomy (Journal for the History of Astronomy Supplement)*, 16(9), S69–S85.

Zeilik, M., 1985b. Sun shrines and Sun symbols in the U.S. Southwest. *Archaeoastronomy (Journal for the History of Astronomy Supplement)*, 16(9), S86–S96.

Zeilik, M., 1985c. The Fajada Butte solar marker: a reevaluation. *Science*, 228, 1311–1313.

Zeilik, M., 1989. Keeping the sacred and planting calendar: archaeoastronomy in the Pueblo Southwest. In Aveni, A.F. (ed.). *World Archaeoastronomy*. Cambridge, Cambridge University Press. Pp. 143–166.

Zoll, K.J., 2010. Prehistoric astronomy of Central Arizona. *Archaeoastronomy*, 23, 154–164.

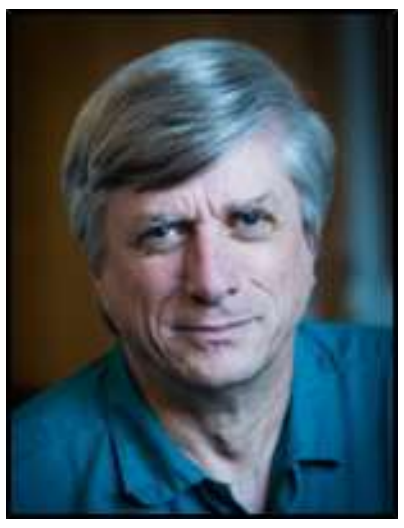

**Dr Bradley E. Schaefer** is the Distinguished Professor in the Department of Physics and Astronomy at the Louisiana State University in Baton Rouge, Louisiana, USA. He is a winner of the 2007 Gruber Prize for Cosmology and the 2015 Breakthrough Prize in Fundamental Physics, as one of the discoverers of the still-enigmatic Dark Energy and the accelerating Universe. With 230 publications in refereed journals, his astrophysics specialty has been variable stars of many types, concentrating on supernovae and novae. Seventy of these publications are on the history of astronomy, with a unique specialization in celestial visibility, in what constitutes a separate career.

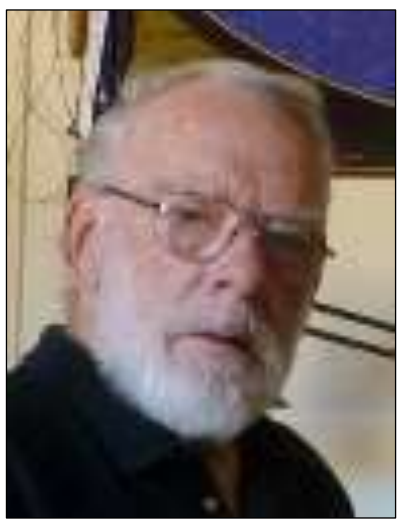

**James Stamm** is now retired, but still pursuing research on a variety of topics related to rocks and rock art in the region around his home in Tucson. These include the sun daggers (with on-going work for eight sun daggers at a nearby site), gong rocks, and labyrinth designs. Since 1983 and still ongoing, Stamm has produced a unique database of the visibility aspects of thin (crescent) moons. Further, Stamm has helped with logistics for the International Occultation Timing Association (IOTA). asteroid occultation program, for observing. IOTA celebrated Stamm's life work with its 2019 David E. Laird Award.